\documentclass[11pt,draftcls,onecolumn]{IEEEtran}
\IEEEoverridecommandlockouts                              
\usepackage{amsmath}
\usepackage{color}
\usepackage{graphicx}
\usepackage{latexsym}

\usepackage{epsfig} 

\title{Asymptotic Behavior of the Pseudo-Covariance Matrix of a Robust State Estimator with Intermittent Measurements}

\author{Tong Zhou
\thanks{This work was supported in part by the
973 Program under Grant 2009CB320602, the National Natural Science
Foundation of China under Grant 61174122 and 61021063, and the
Specialized Research Fund for the Doctoral Program of Higher
Education, P.R.C., under Grant 20110002110045.}
\thanks{T.Zhou is with the Department of Automation and TNList, Tsinghua University, Beijing, 100084,
CHINA. (Tel: 86-10-62797430; Fax: 86-10-62786911; e-mail:
tzhou@mail.tsinghua.edu.cn.)} }

\begin{document}
\renewcommand{\thefootnote}{\fnsymbol{footnote}}
\maketitle 
\renewcommand{\thepage}{33--\arabic{page}}
\setcounter{page}{1}

\begin{abstract}
Ergodic properties and asymptotic stationarity are investigated in this paper for the pseudo-covariance matrix (PCM) of a
recursive state estimator which is robust against parametric uncertainties and is based on plant output measurements that may be randomly dropped.
When the measurement dropping process is described by a Markov chain and the modified plant is both controllable and observable, it is proved that if the dropping probability is less than 1, this PCM converges to a stationary distribution that is independent of its initial values. A convergence rate is also provided. In addition, it has also been made clear that when the initial value of the PCM is set to the stabilizing solution of the algebraic Riccati equation related to the robust state estimator without measurement dropping, this PCM converges to an ergodic process. Based on these results, two approximations are derived for the probability distribution function of the stationary PCM, as well as a bound of approximation errors. A numerical example is provided to illustrate the obtained theoretical results.

\end{abstract}

\IEEEpeerreviewmaketitle

\section{Introduction}

With the development of network technologies, numerous novel anticipations, as well as various new technical issues, rise in system analysis and synthesis, due to the significant differences in information exchange methods between a traditional system and a network system. Among them, one important issue is state estimation with random measurement droppings, in which plant output measurements are stochastically lost due to failures of information delivery from the plant output measurement sensors to its state estimator \cite{ln11,ssfpjs04,km12,mn12}.

Over the last decade, this problem has attracted extensive attentions and various results have been obtained. In \cite{ssfpjs04}, optimality of the traditional Kalman filter is established under the existence of random measurement droppings, provided that information is available in the received data on whether or not it is a measured plant output. It has also been made clear that for an unstable plant, to guarantee boundedness of the
expectation of the covariance matrix of estimation errors, in addition to controllability and observability, the probability that the estimator receives plant output measurements must be higher than some threshold values. Afterwards, it is observed that although simultaneous loss of plant output measurements at all sample instants usually has an essential zero probability to occur, it is the dominating fact that leads to an infinite expectation of this covariance matrix.
This observation results in the importance recognition about the
probability distribution of this covariance matrix which is argued to be a more appropriate measure on the performances of a state estimator with random data missing \cite{Censi11,rmf11,sem10,km12,ms12}.

Particularly, some upper and lower bounds are derived respectively in \cite{sem10,rmf11} for the probability that this covariance matrix
is smaller than a prescribed positive definite matrix (PDM). Under the condition that an unstable plant has a
diagonalizable state transition matrix, \cite{ms12} shows that if some controllability and
observability conditions are satisfied, the trace of this covariance matrix decays according to
a power law. Based on the contractive properties of Riccati recursions and convergence conditions on random
iterated functions, this covariance matrix is proved in \cite{Censi11} to converge in general to a stationary distribution that is independent of its initial values, no matter the measurement loss process is described by a Bernoulli process, a Markov
chain or a semi-Markov chain. When the observation arrival is modeled by a Bernoulli process and the
packet arrival probability is approximately equal to 1, this covariance matrix is shown in \cite{km12} to
converge weakly to a unique invariant distribution satisfying a
moderate deviation principle with a good rate function.

When a plant model is not accurate, which is the general situation in actual engineering applications of a state estimator, recursive state estimations that are robust against modelling errors have also been extensively investigated \cite{george12,ln11,ksh00,nbt07,simon06,zhou10b,zhou11}. Some of these methods have already been extended to systems with an imperfect communication channel, for example, \cite{mn12, zhou14} and the references therein. Especially, in \cite{zhou14}, a robust state estimator is derived using penalizations on the sensitivity of the innovation process of an estimator to parametric modelling errors, which has a similar form as that of the Kalman filter and can be recursively realized without any condition validations and on line design parameter adjustments. Moreover, some necessary and sufficient conditions have also been established on the convergence of the pseudo-covariance matrix (PCM) of this robust state estimator to a stationary distribution, which include the results on Kalman filtering with intermittent observations as special cases.

These investigations have made many important theoretical issues clear
about state estimations with random measurement arrivals, and the obtained results appear greatly helpful in the analysis and synthesis of networked
systems. Some important issues of this state estimation problem, however, still need further efforts. Among them, one essential problem is about a more accurate characterization of the stationary distribution of the covariance matrix in the Kalman filtering or the PCM in the robust estimations, as this characterization is directly connected with their estimation performances and is important in determining requirements on the communication channel \cite{Censi11,km12,zhou14}.

This paper discusses properties of the stationary distribution of the PCM in the sensitivity
penalization based robust state estimations with random measurement droppings.
The data missing process is assumed to be described by a Markov chain, which can include the Bernoulli process as a special case.
On the basis of a Riemannian metric on the space of positive definite matrices (PDM) and a central limit theorem for Markov chains, it is proved that when the modified plant in the robust estimations is both controllable and observable, this PCM converges to a stationary distribution, provided that the data arrival probability is greater than zero. A convergence rate is also given. It has also been shown that when the PCM is started from the stablilizing solution to the algebraic Riccati equation defined by a modified plant, the PCM process is both stationary and ergodic. From these results, two approximations are given for the stationary distribution of the PCM with an arbitrary Markov chain probability transition matrix, as well as its convergence rate to the actual value. These results are also valid for the covariance matrix of the Kalman filter with intermittent observations.

The outline of this paper is as follows. At first, in Section II,
the sensitivity penalization based robust state estimation procedure with intermittent observations is briefly summarized, and some preliminary results on Markov process and Riccati recursions are provided.  Afterwards, stationarity and ergodicity properties of the PCM process are investigated in Section III, while Section IV derives an approximation of the stationary distribution of the PCM, as well as its convergence rate to the actual value. A numerical example is provided in Section V
to illustrate the effectiveness and accuracy of the suggested approximation method. Finally,
some concluding remarks are given in Section VI. An appendix is included to
give proofs of some technical results.

The following notation and symbols are adopted.
The product $\Phi_{k1}\Phi_{k1-1\;{\rm or}\; k1+1}\cdots\Phi_{k2}$ is denoted by
$\prod_{j=k1}^{k2}\Phi_{j}$, while the transpose of a matrix/vector is indicated by the superscript $T$. For matrices $P$ and
$\Phi=\left[\left.\Phi_{ij}\right|_{i,j=1}^{2}\right]$ with compatible dimensions, a Homographic transformation ${\rm\bf
H}_{m}(\Phi,\;P)$ is defined as ${\rm\bf
H}_{m}(\Phi,\;P)=[\Phi_{11}P+\Phi_{12}][\Phi_{21}P+\Phi_{22}]^{-1}$.
${\rm\bf P}_{r}(\cdot)$ is used to denote the probability of the
occurrence of a random event, while ${\rm\bf
E}_{\{\sharp\}}\!\{\star\}$ and ${\rm\bf
V}_{{ ar}\{\sharp\}}\!\{\star\}$ the mathematical expectation of a random matrix
valued function (MVF) $\star$ with respect to the random variable
$\sharp$ and the variance of a random variable $\star$. The subscript $\sharp$ is usually omitted when it is
obvious. $O(x)$ stands for a number that is of the same order in magnitude as $x$, while $\Phi(t)$ the distribution function of a normally distributed random variable with its mathematical expectation and variance respectively being $0$ and $1$. $I_{\cal A}(x)$ is the indictor function which equals $1$ when $x$ belongs to the set $\cal A$ and zero elsewhere, and $\#\{\star\}$ the number of elements in a set.

\section{The Robust State Estimation Procedure and Some Preliminaries}

Assume that the input output relations of a linear time varying dynamic system $\rm\bf\Sigma$ can
be described by the following discrete state-space model,
\begin{equation}
{\rm\bf\Sigma}: \hspace{0.5cm}\left\{\begin{array}{l}
x_{k+1}=A_{k}(\varepsilon_{k})x_{k}+B_{k}(\varepsilon_{k})w_{k} \\
y_{k}=\gamma_{k}C_{k}(\varepsilon_{k})x_{k}+v_{k} \end{array}
\right. \label{eqn:1}
\end{equation}
in which vectors $w_{k}$ and $v_{k}$ denote respectively process noises and composite
influences of measurement errors and communication errors, the $n_{e}$ dimensional vector $\varepsilon_{k}$ stands for
plant parametric errors at the time instant
$k$, while the  random variable $\gamma_{k}$ describes characteristics of the communication channel from the plant output measurement sensor to its state estimator. It takes a value from the set $\{\:0,\;1\:\}$ which respectively represents that a plant output measurement is successfully transmitted or the communication channel is out of order. An assumption adopted throughout this paper is that this random variable $\gamma_{k}$ is a Markov chain with its probability transitions described by
\begin{equation}
\left[\begin{array}{c} {\rm\bf P}_{r}(\gamma_{k}=1) \\  {\rm\bf P}_{r}(\gamma_{k}=0) \end{array}\right]
=\left[\begin{array}{cc} \alpha_{k} & 1-\beta_{k} \\ 1-\alpha_{k} & \beta_{k} \end{array}\right]
\left[\begin{array}{c} {\rm\bf P}_{r}(\gamma_{k-1}=1) \\  {\rm\bf P}_{r}(\gamma_{k-1}=0) \end{array}\right]
\end{equation}
in which $\alpha_{k}$ and $\beta_{k}$ are two deterministic functions of the temporal variable $k$ and take values only from the interval $(0,\;1)$. This model is widely adopted in the description of a communication channel, and is sometimes called the Gilbert-Elliot model \cite{Censi11,ms12,sem10}. It is also assumed throughout this paper that the state vector $x_{k}$ of the dynamic system $\rm\bf\Sigma$ has a dimension $n$, and an indicator
is included in the received signal $y_{k}$ that reveals whether or not it contains information about plant outputs.

In \cite{zhou14}, it is assumed that both $w_{k}$ and $v_{k}$ are white and
normally distributed with ${\rm\bf E}\!\left({\rm\bf
col}\!\{w_{k},v_{k},x_{0}\}\right)=0$ and ${\rm\bf E}\!\left({\rm\bf
col}\!\{w_{k}, v_{k},x_{0}\}{\rm\bf col}^{T}\!\{w_{s},
v_{s},x_{0}\}\right)={\rm\bf diag}\{Q_{k}\delta_{ks},
R_{k}\delta_{ks},P_{0}\}$, $\forall k,s>0$, in which $\delta_{ks}$
stands for the Kronecker delta function, and
$Q_{k}$ and $R_{k}$ are known positive definite MVFs of the temporal
variable $t$, while $P_{0}$ is a known PDM. Another
hypothesis adopted in \cite{zhou14} is that all the system matrices
$A_{k}(\varepsilon_{k})$, $B_{k}(\varepsilon_{k})$ and
$C_{k}(\varepsilon_{k})$  are time varying but
known MVFs with each of their elements differentiable with respect to every
element of the modelling error vector $\varepsilon_{k}$ at each time instant. Under these assumptions, the following recursive robust state estimator is derived in \cite{zhou14} for the system $\rm\bf\Sigma$, which is abbreviated as RSEIO.

\hspace*{-0.4cm}{\bf State Estimation Procedure (RSEIO).} Let $\mu_{k}$ denote the positive design parameter belonging to
$(0,\;1]$ that reflects a trade-off between nominal value of
estimation accuracy and its sensitivities to
parametric modelling errors. Define $\lambda_{k}$ as
$\lambda_{k}=\frac{1-\mu_{k}}{\mu_{k}}$. Assume that both $P_{k|k}$ and $Q_{k}$
are invertible, in which $P_{k|k}$ is the PCM of the state estimator at the time instant $k$. It is proved in \cite{zhou14} that the estimate of the state vector $x_{k+1}$ of
the dynamic system $\rm\bf\Sigma$ based on $y_{k}|_{k=0}^{t+1}$ has the following
recursive expression,
\begin{equation}
\hat{x}_{k+1|k+1}=\left\{\begin{array}{ll} A_{k}(0)\hat{x}_{k|k} &
\gamma_{k+1}=0 \\
\hat{A}_{k}(0)\hat{x}_{k|k}+P_{k+1|k+1}C_{k+1}^{T}(0)R_{k+1}^{-1}\{y_{k+1}-C_{k+1}(0)\hat{A}_{k}(0)\hat{x}_{k|k}\}
& \gamma_{k+1}=1
\end{array}\right.
\end{equation}
Moreover, the PCM $P_{k|k}$ can be recursively updated as
\begin{equation}
P_{k+1|k+1}\!=\!\left\{\!\!\begin{array}{ll}
A_{k}(0)P_{k|k}A_{k}^{T}(0)+B_{k}(0)Q_{k}B_{k}^{T}(0) &
\gamma_{k+1}=0 \\
\left\{\left[A_{k}(0)\hat{P}_{k|k}A_{k}^{T}(0)+\hat{B}_{k}(0)\hat{Q}_{k}\hat{B}_{k}^{T}(0)\right]^{-1}+C_{k+1}^{T}(0)R_{k+1}^{-1}C_{k+1}(0)\right\}^{-1}
& \gamma_{k+1}=1
\end{array}\right.
\label{eqn:5}
\end{equation}
in which
\begin{eqnarray*}
& & \hspace*{-1cm}\hat{P}_{k|k}=(P_{k|k}^{-1}+\lambda_{k}S_{k}^{T}S_{k})^{-1}, \hspace{0.5cm}
\hat{Q}_{k}=\left[Q_{k}^{-1}+\lambda_{k}T_{k}^{T}(I+\lambda_{k}S_{k}P_{k|k}S_{k}^{T})T_{k}\right]^{-1} \\
& & \hspace*{-1cm}
\hat{B}_{k}(0)=B_{k}(0)-\lambda_{k}A_{k}(0)\hat{P}_{k|k}S_{k}^{T}T_{k},\hspace{0.5cm}
\hat{A}_{k}(0)=[A_{k}(0)-\hat{B}_{k}(0)\hat{Q}_{k}T_{k}^{T}S_{k}][I-\lambda_{k}\hat{P}_{k|k}S_{k}^{T}S_{k}]\\
& & \hspace*{-1cm}S_{k}={\rm\bf col}\!\left.\left\{\!\!\left[\!\begin{array}{cc}
C_{k+1}(\varepsilon_{k+1})\frac{\partial(A_{k}(\varepsilon_{k}))}{\partial\varepsilon_{k,k}}
\\
\frac{\partial(C_{k+1}(\varepsilon_{k+1}))}{\partial\varepsilon_{k+1,k}}A_{k}(\varepsilon_{k})
\end{array}\!\right]_{k=1}^{n_{e}}\!\!\right\}\right|\!\!\!{\footnotesize\begin{array}{l}
\\ \varepsilon_{k}=0 \vspace{-0.25cm}
\\ \varepsilon_{k+1}=0\end{array}}\!\!\!\!\!,\hspace{0.25cm}
T_{k}={\rm\bf col}\!\left.\left\{\!\!\left[\!\begin{array}{cc}
C_{k+1}(\varepsilon_{k+1})\frac{\partial(B_{k}(\varepsilon_{k}))}{\partial\varepsilon_{k,k}}
\\
\frac{\partial(C_{k+1}(\varepsilon_{k+1}))}{\partial\varepsilon_{k+1,k}}B_{k}(\varepsilon_{k})
\end{array}\!\right]_{k=1}^{n_{e}}\!\!\right\}\right|\!\!\!{\footnotesize\begin{array}{l}
\\ \varepsilon_{k}=0 \vspace{-0.25cm}
\\ \varepsilon_{k+1}=0\end{array}}
\end{eqnarray*}

When $S_{k}\equiv 0$ and $T_{k}\equiv 0$, the above recursive state estimation procedure reduces to the Kalman filter with intermittent observations \cite{zhou14}. As the results of this paper depend neither on $S_{k}$ nor on $T_{k}$, it can be claimed that they are also valid for the Kalman filtering with random dada droppings.

Concerning this state estimation procedure, it has also been proved in \cite{zhou14} that if the matrix
$A_{k}(0)-\lambda_{k}B_{k}(0)(Q_{k}^{-1}+\lambda_{k}T_{k}^{T}T_{k})^{-1}T_{k}^{T}S_{k}$,
denote it by $\check{A}_{k}$, is invertible, then, the PCM $P_{k+1|k+1}$ with $\gamma_{k+1}\neq 0$ can be more compactly expressed as
\begin{equation}
P_{k+1|k+1}^{-1}=\left[\tilde{A}_{k}P_{k|k}\tilde{A}_{k}^{T}+B_{k}(0)\tilde{Q}_{k}B_{k}^{T}(0)\right]^{-1}+\tilde{C}_{k+1}^{T}\tilde{R}^{-1}_{k+1}\tilde{C}_{k+1}
\end{equation}
in which matrices
$\tilde{A}_{k}$, $\tilde{B}_{k}$, $\tilde{C}_{k+1}$, $\tilde{Q}_{k}$, $\check{Q}_{k}$
and $\tilde{R}_{k+1}$ respectively have the following definitions,
\begin{eqnarray*}
& &
\tilde{A}_{k}=\check{A}_{k}+B_{k}(0)\check{Q}_{k}\tilde{B}_{k}^{T}\tilde{S}_{k}^{T}\tilde{S}_{k},\hspace{0.25cm}
\tilde{B}_{k}=\check{A}_{k}^{-1}B_{k}(0),\hspace{0.25cm}\tilde{Q}_{k}=\check{Q}_{k}
+\check{Q}_{k}\tilde{B}_{k}^{T}\tilde{S}_{k}^{T}\tilde{S}_{k}\tilde{B}_{k}\check{Q}_{k}\\
& & \check{Q}_{k}=(Q_{k}^{-1}+\lambda_{k}T_{k}^{T}T_{k})^{-1}, \hspace{0.25cm}
\tilde{S}_{k}\!=\!\sqrt{\lambda_{k}}\left[I+\lambda_{k}T_{k}Q_{k}T_{k}^{T}\right]^{-1/2}S_{k}\\
& &
\tilde{C}_{k+1}\!=\!\left[\begin{array}{c}
\tilde{S}_{k}\check{A}_{k}^{-1} \\ C_{k+1}(0)
\end{array}\right],\hspace{0.25cm}
\tilde{R}_{k+1}\!=\!\left[\begin{array}{cc}
I+\tilde{S}_{k}\tilde{B}_{k}\check{Q}_{k}\tilde{B}_{k}\tilde{S}_{k}^{T}
& 0 \\ 0 & R_{k+1} \end{array}\right]
\end{eqnarray*}
While this expression for $P_{k+1|k+1}$ is much more complicated than that of Equation (\ref{eqn:5}), it is more convenient in analyzing properties of the robust state estimator, as it gives a relation of the PCMs of RSEIO at two successive time instants.

In studying asymptotic properties of Riccati recursions, an efficient metric is a Riemannian distance between two PDMs \cite{Bougerol93,Censi11,zhou14}. More precisely, let $P$ and $Q$ be two $n\times n$ dimensional PDMs and
$\lambda_{i}$ an eigenvalue of the matrix $PQ^{-1}$. Then, the Riemannian distance between these two matrices, denote it by
$\delta(P,Q)$, is defined as $\delta(P,Q)=\sqrt{\sum_{i=1}^{n}{
ln}^{2}(\lambda_{i})}$. In combination with properties of Hamiltonian matrices and Homographic transformations, this metric plays an essential role in the following analysis on the asymptotic properties of RSEIO.

To analyze asymptotic properties of the PCM $P_{k|k}$, it is
assumed throughout this paper that the nominal model of
the plant, as well as the first order derivatives at the origin of the innovation
process $e_{k}(\varepsilon_{k},\varepsilon_{k+1})$ with respect to
every parametric modelling error, that is, the matrices $S_{k}$ and $T_{k}$, do not change with the temporal variable
$k$. Under such a situation, it is feasible to define temporal variable $k$ independent
matrices $A^{[1]}$, $A^{[0]}$, $G^{[1]}$, $G^{[0]}$ and $H^{[1]}$ respectively as
\begin{displaymath}
A^{[1]}=\tilde{A}_{k},
\hspace{0.5cm}G^{[1]}=B_{k}{\tilde{Q}_{k}^{1/2}},\hspace{0.5cm}
H^{[1]}=\tilde{R}_{k+1}^{-1/2}\tilde{C}_{k+1},
 \hspace{0.5cm} A^{[0]}=A_{k},
\hspace{0.5cm} G^{[0]}=B_{k}Q_{k}^{1/2}
\end{displaymath}
Assume that both $A^{[0]}$ and $A^{[1]}$ are invertible. Using these matrices, define matrices $M^{[0]}$ and $M^{[1]}$ respectively as
\begin{displaymath}
M^{[0]}=\left[\begin{array}{cc} A^{[0]} &
G^{[0]}G^{[0]T}A^{[0]-T} \\ 0 & \left(A^{[0]}\right)^{-T} \end{array}\right],\hspace{0.5cm}
M^{[1]}=\left[\begin{array}{cc} A^{[1]} & G^{[1]}G^{[1]T}\left(A^{[1]}\right)^{-T} \\
H^{[1]T}H^{[1]}A^{[1]}
&
[I+H^{[1]T}H^{[1]}G^{[1]}G^{[1]T}]\left(A^{[1]}\right)^{-T}\end{array}\right]
\end{displaymath}
Then, according the results of \cite{zhou14}, both $M^{[0]}$ and $M^{[1]}$ are Hamiltonian and the recursion for the PCM of the RSEIO can be reexpressed as
\begin{equation}
P_{k+1|k+1}=\left\{\begin{array}{lll} {\rm\bf H}_{m}(M^{[0]},\;P_{k|k}) & \;\;\;\;\; &
\gamma_{k+1}=0 \\
{\rm\bf H}_{m}(M^{[1]},\;P_{k|k}) & \;\;\;\;\; & \gamma_{k+1}=1 \end{array}\right.
\label{eqn:2}
\end{equation}
Moreover, ${\rm\bf H}_{m}(M^{[0]},X)$ and ${\rm\bf H}_{m}(M^{[1]},X)$ are always well defined whenever the matrix $X$ is a PDM with a compatible dimension. Furthermore, when $P_{0|0}$ is positive definite which is generally satisfied in practical engineering problems, the following relation exists between the PCM $P_{k|k}$ and its initial value $P_{0|0}$,
\begin{equation}
P_{k|k}={\rm\bf H}_{m}\left(M^{[\gamma_{k}]},\;{\rm\bf
H}_{m}\left(M^{[\gamma_{k-1}]},\;\cdots,\;{\rm\bf
H}_{m}\left(M^{[\gamma_{1}]},\;P_{0|0}\right)\cdots\right)\right)
={\rm\bf H}_{m}\left(\prod_{i=k}^{1}M^{[\gamma_{i}]},\;P_{0|0}\right)
\label{eqn:6}
\end{equation}

To analyze asymptotic properties of the PCM of the robust state estimator RSEIO, the following results on Markov process are also
needed.

\hspace*{-0.4cm}{\bf Lemma 1.}\cite{lr76, stenflo12} Let $x_{i}|_{i=0}^{\infty}$ be a positive recurrent irreducible Markov chain defined by a probability space $(\Omega,{\cal F},P)$ with a countable state space $\cal I$, and $f(\cdot)$ be a real valued function defined on $\cal I$. Denote the $\alpha$-th entrance of the Markov chain into its $j$-th state by $\tau_{\alpha}^{[j]}$, and $\sum_{k=\tau_{\alpha}^{[j]}}^{\tau_{\alpha+1}^{[j]}-1}f(x_{k})$ by $f_{\alpha}^{[j]}$. If both ${\rm\bf E}\left(|f_{\alpha}^{[j]}|^{3}\right)$ and
${\rm\bf E}\left(|\tau_{\alpha+1}^{[j]}-\tau_{\alpha}^{[j]}|^{3}\right)$ are finite, and $\sigma_{j}=\sqrt{{\rm\bf V}_{ar}\{f_{\alpha}^{[j]}-s(f)(\tau_{\alpha+1}^{[j]}-\tau_{\alpha}^{[j]})\}}$ is greater than $0$, then,
\begin{equation}
\sup_{t\in{\cal R}}\left|{\rm\bf P}_{r}\left\{\;\frac{1}{\sigma_{j}\sqrt{n\pi_{j}}}\left(\sum_{k=0}^{n}f(x_{k})-(n+1)s(f)\right)<t\right\}-\Phi(t)\right|=O\left(\left(\frac{ln(n)}{n}\right)^{1/4}\right)
\end{equation}
in which $s(f)=\sum_{i\in{\cal I}}\frac{f(i)}{\mu_{i}}$ with $\mu_{i}$ the mathematical expectation of the recurrence time of the $i$-th state, and $\pi_{i}=\mu_{i}^{-1}$.

\hspace*{-0.4cm}{\bf Lemma 2.}\cite{elton87} Assume that a Markov process $x_{i}|_{i=0}^{\infty}$ has an unique stationary distribution $\mu$. Then, this process with $x_{0}$ having distribution $\mu$ is ergodic.

\section{Asymptotic Properties of the PCM}

From the state estimation procedure, it is clear that all the asymptotic properties of the RSEIO are dominated by those of the PCM, which is very similar to that of the Kalman filter, although in which the covariance matrix has a more clear physical interpretation and is more closely related to its estimation accuracies. In this section, the preliminary results given in the previous section are utilized to establish asymptotic behaviours of the PCM of the robust state estimator RSEIO, under the condition that both the nominal plant model parameters and the sensitivity of the innovation process to parametric modelling errors are time invariant. To simplify expressions, the subscripts for $\alpha_{k}$ and $\beta_{k}$ are omitted, and the system with its state space model parameters being $(A^{[1]},\;
G^{[1]},\;H^{[1]})$ is called the modified plant.

Major results of this section include stationarity and ergodicity of the random PCM process. More precisely, it is at first proved that for arbitrary $0<\alpha,\;\beta<1$, if the modified plant is both controllable and observable, then, the PCM of the RSEIO converges in an exponential rate to a stationary process independent of its initial values. Moreover, if the initial value of the PCM takes the value of the stablilizing solution of the algebraic Riccati equation defined by the Kalman filter for the modified plant, then, the random process PCM is also ergodic. These results are also valid for Kalman filtering with intermittent observations, noting that when there are no modelling errors in the system $\rm\bf\Sigma$, the robust state estimator RSEIO reduces to the Kalman filter.

To establish these properties, the following symbols are introduced.
\begin{eqnarray*}
& & \Phi_{k}(X)={\rm\bf H}_{m}\left(M^{[\gamma_{k}]},\;{\rm\bf
H}_{m}\left(M^{[\gamma_{k-1}]},\;\cdots,\;{\rm\bf
H}_{m}\left(M^{[\gamma_{1}]},\;X\right)\cdots\right)\right),\hspace{0.5cm} \gamma_{i}\in\{0,\;1\} \\
& & \delta_{k}(X,Y)=\delta(\Phi_{k}(X),\Phi_{k}(Y))
\end{eqnarray*}

\hspace*{-0.4cm}{\bf Theorem 1.} Assume that the modified plant $(A^{[1]},\;
G^{[1]},\;H^{[1]})$ is both controllable and observable. Then, for arbitrary $\alpha,\;\beta$ belonging to the open interval $(0,\;1)$ and arbitrary PDMs $X$ and $Y$,
\begin{equation}
\lim_{n\rightarrow\infty}\delta_{n}(X,Y)=0,\hspace{0.5cm} {\rm in \;\; probability}
\end{equation}

A proof of this theorem is given in the appendix.

Theorem 1 and Equation (\ref{eqn:6}) make it clear that if the matrix pair $(A^{[1]},\;
G^{[1]})$ is controllable and the matrix pair $(H^{[1]},\;A^{[1]})$ is observable, and the Markov chain $\gamma_{k}$ does not degenerate into two isolated states, then, the limit PCM $P_{\infty|\infty}$ of the RSEIO is independent of its initial value $P_{0|0}$. Moreover, from Equation (\ref{eqn:a10}), it can be understood that from any initial value, the convergence of the PCM $P_{k|k}$ to its limit $P_{\infty|\infty}$ is exponential.

Define a set $\cal P$ as
\begin{equation}
{\cal P}=\left\{\:P\:\left|\:P=\lim_{n\rightarrow\infty}{\rm\bf H}_{m}\left(\prod_{i=n}^{1}M^{[\gamma_{i}]},\;X\right),\;\;X>0,\;\gamma_{i}\in\{0,\;1\}\right.\right\}
\label{eqn:7}
\end{equation}
Then, Theorem 1 makes it clear that when the adopted assumptions are satisfied, this matrix set is independent of a particular PDM $X$. On the other hand, from its definition, it is obvious that this matrix set consists of all the final value of the PCM of the RSEIO.

For an arbitrary $P\in{\cal P}$, there exists a corresponding series $\gamma_{i}|_{i=1}^{\infty}$, such that $P=\lim_{n\rightarrow\infty}{\rm\bf H}_{m}\left(\prod_{i=n}^{1}M^{[\gamma_{i}]},\right.$ $\left. X\right)$. Therefore, for every $\gamma\in\{0,\;1\}$,
\begin{eqnarray}
{\rm\bf H}_{m}\left(M^{[\gamma]},\;P\right)&=&{\rm\bf H}_{m}\left[M^{[\gamma]},\;\lim_{n\rightarrow\infty}{\rm\bf H}_{m}\left(\prod_{i=n}^{1}M^{[\gamma_{i}]},\;X\right)\right]\nonumber\\
&=&{\rm\bf H}_{m}\left(\lim_{n\rightarrow\infty}M^{[\gamma]}M^{[\gamma_{n}]}M^{[\gamma_{n-1}]}\cdots
M^{[\gamma_{1}]},\;X\right)\nonumber\\
&=&{\rm\bf H}_{m}\left[\lim_{n\rightarrow\infty}M^{[\gamma]}M^{[\gamma_{n}]}M^{[\gamma_{n-1}]}\cdots
M^{[\gamma_{2}]},\;{\rm\bf H}_{m}\left(M^{[\gamma_{1}]},\;X\right)\right]\nonumber\\
&=&{\rm\bf H}_{m}\left(\lim_{n\rightarrow\infty}M^{[\gamma]}M^{[\gamma_{n}]}M^{[\gamma_{n-1}]}\cdots
M^{[\gamma_{2}]},\;X\right)\hspace{0.5cm} {\rm (\;in\;\;probability\;)}\nonumber\\
&\in& {\cal P}
\end{eqnarray}

On the contrary, let $\gamma=\gamma_{n}\in\{0,\;1\}$. Then,
\begin{eqnarray}
P&=&\lim_{n\rightarrow\infty}{\rm\bf H}_{m}\left(\prod_{i=n}^{1}M^{[\gamma_{i}]},\;X\right)\nonumber\\
&=&\lim_{n\rightarrow\infty}{\rm\bf H}_{m}\left[M^{[\gamma_{n}]},\;{\rm\bf H}_{m}\left(\prod_{i=n-1}^{1}M^{[\gamma_{i}]},\;X\right)\right]\nonumber\\
&=&\lim_{n\rightarrow\infty}{\rm\bf H}_{m}\left[M^{[\gamma]},\;{\rm\bf H}_{m}\left(\prod_{i=n-1}^{1}M^{[\gamma_{i}]},\;X\right)\right]\nonumber\\
&=&{\rm\bf H}_{m}\left[M^{[\gamma]},\;\lim_{n\rightarrow\infty}{\rm\bf H}_{m}\left(\prod_{i=n-1}^{1}M^{[\gamma_{i}]},\;X\right)\right]
\end{eqnarray}
Obviously from the definition of the set $\cal P$, $\lim_{n\rightarrow\infty}{\rm\bf H}_{m}\left(\prod_{i=n-1}^{1}M^{[\gamma_{i}]},\;X\right)\in{\cal P}$.
This means that there exists at least one $\bar{P}\in{\cal P}$, such that $P={\rm\bf H}_{m}\left(M^{[\gamma]},\;\bar{P}\right)$.

On the basis of these relations, it seems very possible that when the conditions of Theorem 1 are satisfied, two successive random PCMs, say $P_{k|k}$ and $P_{k+1|k+1}$, have the same support when the temporal variable $k$ is large. This imply that the final value of the PCM of the robust state estimator RSEIO, that is, $P_{\infty|\infty}$, may have a unique stationary distribution. As a matter of fact, this stationarity can be declared from Theorem 5 of \cite{zhou14}.

When $(A^{[1]},\; G^{[1]})$ is controllable and $(H^{[1]},\;A^{[1]})$ is observable, a well established conclusion in control theory is that the following algebraic Riccati equation
\begin{equation}
P=\left[(A^{[1]}PA^{[1]T}+G^{[1]}G^{[1]T})^{-1}+H^{[1]T}H^{[1]}\right]^{-1}
\end{equation}
has a unique stabilizing solution. This stabilizing solution is denoted by $P^{\star}$ throughout the rest of this paper. Moreover, a widely known result in Kalman filtering is that under these conditions, the Riccati recursion
$P_{k+1|k+1}=\left[(A^{[1]}P_{k|k}A^{[1]T}+G^{[1]}G^{[1]T})^{-1}+H^{[1]T}H^{[1]}\right]^{-1}$ converges to $P^{\star}$ with the increment of the temporal variable $k$ \cite{ksh00,simon06}.

On the basis of these results, ergodicity of the random PCM process is established.

\hspace*{-0.4cm}{\bf Corollary 1.} In addition to the conditions of Theorem 1, if the PCM of the robust state estimator RSEIO starts from $P^{\star}$, then, this random process is also ergodic.

\hspace*{-0.4cm}{\bf Proof:} When the conditions of Theorem 1 are satisfied, from Theorem 5 of \cite{zhou14}, it can be claimed that the PCM of the RSEIO converges to a stationary distribution. Theorem 1 makes it clear that this stationary distribution is unique and the convergence rate is exponential.

On the other hand, if $\gamma_{k}\equiv 1$, $k=1,2,\cdots$, then, for an arbitrary PDM $X$,
\begin{equation}
P_{\infty|\infty}=\lim_{k\rightarrow\infty}{\rm\bf H}_{m}\left(M^{[1]k},\; X\right)
\end{equation}
When $(A^{[1]},\; G^{[1]})$ is controllable and $(H^{[1]},\;A^{[1]})$ is observable, from the convergence properties of the Kalman filter \cite{ksh00,simon06}, we have that $\lim_{k\rightarrow\infty}{\rm\bf H}_{m}\left(M^{[1]k},\;X\right)=P^{\star}$. Moreover, from the definition of the matrix $P^{\star}$, it is obvious that ${\rm\bf H}_{m}\left(M^{[1]},\;P^{\star}\right)=P^{\star}$. Therefore, $P^{\star}$ belongs to the support of the stationary distribution of the random process $P_{k|k}$.

It can therefore be declared from Lemma 2 that the random process $P_{k|k}$ initialized with $P_{0|0}=P^{\star}$ is ergodic.

This completes the proof. \hspace{\fill}$\Diamond$

When both $\alpha$ and $\beta$ belong to the open set $(0,\;1)$, it can be directly proved, as what has been done in \cite{mt93}, that the Markov chain $\gamma_{k}|_{k=1}^{\infty}$ has a stationary distribution. Denote the random variable of this stationary distribution by $\gamma$. Then, at its stationary state, the probability that $\gamma_{k}$ takes the value of $1$ or $0$ does not depend on the temporal variable $k$, which can be respectively expressed as ${\rm\bf P}_{r}(\gamma=1)=\frac{1-\beta}{2-\alpha-\beta}$ and ${\rm\bf P}_{r}(\gamma=0)=\frac{1-\alpha}{2-\alpha-\beta}$.

From Corollary 1, it is clear that the stationary distribution of the random process $P_{k|k}$ can be approximated well by its time series samples. To clarify accuracy of this approximation, properties of a Markov process are utilized.

For a binary series $\gamma_{i}|_{i=0}^{-\infty}$ with $\gamma_{i}\in\{0,\;1\}$, define $n(\gamma_{i}|_{i=0}^{-\infty})$ and $P^{[n]}$ respectively as $n(\gamma_{i}|_{i=0}^{-\infty})=\sum_{i=0}^{-\infty}\gamma_{i}2^{i}$ and $P^{[n]}=\lim_{k\rightarrow\infty}{\rm\bf H}_{m}\left(M^{[\gamma_{0}]}
M^{[\gamma_{-1}]}\cdots M^{[\gamma_{-k}]},\;P^{\star}\right)$. Moreover, for a prescribed positive number $\varepsilon$, define the set ${\cal P}^{[n]}(\varepsilon)$ of PDMs as
\begin{equation}
{\cal P}^{[n]}(\varepsilon)=\left\{\:P\:\left|\:\delta(P^{[n]},\;P)\leq\varepsilon,\;\;P\geq 0\:\right.\right\}
\end{equation}
Then, according to Theorem 1, for any $n_{1}$ and $n_{2}$ with $n_{\#}=n(\gamma_{i}^{[\#]}|_{i=0}^{\infty})$ and $\#=1,2$, there exists at least one finite length binary sequence $\gamma_{i}^{[n_{1},n_{2}]}|_{i=1}^{N(n_{1},n_{2})}$ with $\gamma_{i}^{[n_{1},n_{2}]}\in\{0,\;1\}$, such that
\begin{equation}
{\rm\bf H}_{m}\left(\prod_{i=N(n_{1},n_{2})}^{1}M^{[\gamma_{i}^{[n_{1},n_{2}]}]},\;P^{[n_{1}]}\right)\in{\cal P}^{[n_{2}]}(\varepsilon) \hspace{0.5cm} {\rm (\;in\;\;probability\;)}
\end{equation}

Note that when $\bar{\gamma}_{i}=\gamma_{i-k}$, we have that $M^{[\bar{\gamma}_{i}]}=M^{[\gamma_{i-k}]}$, $i=0,1,\cdots,n$. This means that
\begin{equation}
{\rm\bf H}_{m}\left(M^{[\gamma_{k}]}
M^{[\gamma_{k-1}]}\cdots M^{[\gamma_{1}]},\;P^{\star}\right)={\rm\bf H}_{m}\left(M^{[\bar{\gamma}_{0}]}
M^{[\bar{\gamma}_{-1}]}\cdots M^{[\bar{\gamma}_{-k}]},\;P^{\star}\right)
\end{equation}
and this relation is valid for all the positive integer (including $+\infty$). It can therefore be declared that the matrix set $\cal P$ defined in Equation (\ref{eqn:7}) can also be expressed as
\begin{equation}
{\cal P}=\left\{\:P^{[n]}\:\left|\: P^{[n]}={\rm\bf H}_{m}\left(\prod_{i=0}^{-\infty}M^{[\gamma_{i}]},\;\; P^{\star}\right),\;\;\right. n=\sum_{i=0}^{-\infty}\gamma_{i}2^{i},\;\;\gamma_{i}\in\{0,\;1\}\:\right\}
\end{equation}
In other words, the set $\cal P$ can be parametrized by $P^{[n]}$ and is therefore countable.

On the other hand, Theorem 1 declares that when $(A^{[1]},\; G^{[1]})$ is controllable and $(H^{[1]},\;A^{[1]})$ is observable and $\alpha,\beta\in(0,\;1)$,
$\lim_{k\rightarrow\infty}\delta_{k}(X,Y)=0$ in probability for arbitrary PDMs $X$ and $Y$. It can therefore be declared that for arbitrary $P^{[p]}$ and $P^{[q]}$ belonging to the set $\cal P$, there exists a binary series $\gamma_{j}^{[pq]}|_{j=1}^{\infty}$ with $\gamma_{j}^{[pq]}\in\{\:0,\;1\:\}$, such that the following equation is valid in probability
\begin{equation}
P^{[p]}=\lim_{k\rightarrow\infty}
{\rm\bf H}_{m}\left(M^{[{\gamma}_{k}^{[pq]}]}
M^{[{\gamma}_{k-1}^{[pq]}]}\cdots M^{[{\gamma}_{1}^{[pq]}]},\;P^{[q]}\right)
\end{equation}
In addition, it has been mentioned before that for an arbitrary positive $\varepsilon$, only finite steps are required in probability to transform an element of ${\cal P}^{[p]}$ to the set ${\cal P}^{[q]}$ by the robust state estimator RSEIO. Note that ${\cal P}^{[p]}$ degenerates into $\{P^{[p]}\}$ when $\varepsilon$ decreases to $0$. This means that the Markov chain $P_{k|k}$ is approximately irreducible and positive recurrent.

Based on these observations, the following results are obtained, whose proof is deferred to the appendix.

\hspace*{-0.4cm}{\bf Theorem 2:} Let $F(x)$ denote the distribution function of the stationary $\delta(P_{\infty|\infty},\;P^{\star})$, and $P_{k|k}$ the PCM of RSEIO at the $k$-th time instant with its initial value $P_{0|0}=P^{\star}$ and the corresponding Markov chain $\gamma_{k}|_{k=0}^{\infty}$ being at its stationary state. For an arbitrary positive number $\varepsilon$, define
${\cal B}_{\varepsilon}$ as ${\cal B}_{\varepsilon}=\left\{\:P\:\left|\:\delta(P,\;P^{\star})\leq \varepsilon\:\right.\right\}$. Then,
\begin{equation}
\lim_{n\rightarrow\infty}\frac{1}{n+1}\sum_{k=0}^{n}I_{{\cal B}_{\varepsilon}}(P_{k|k})=F(\varepsilon),\;\;\;{\rm in\;\; probability}
\end{equation}
and the convergence rate is of order $\left(\frac{ln(n)}{n}\right)^{1/4}$.

From Theorem 2, it can be declared that when the stationary distribution of the random process $P_{k|k}$ is approximated by that of its samples, the approximation accuracy is of order $\left(\frac{ln(n)}{n}\right)^{1/4}$. Therefore, when a large number of the PCM samples $P_{k|k}$ are available, the distribution function of the stationary PCM process can be approximated in a high accuracy.

\section{Approximation of the Stationary Distribution}

In the previous section, it has been proved that when the pseudo-covariance matrix $P_{k|k}$ of the robust state estimator RSEIO starts from $P^{\star}$ and the Markov chain $\gamma_{k}$ is in its stationary state, the corresponding PCM sequence $P_{k|k}$ is ergodic. These results make it possible to approximate the stationary distribution of $P_{k|k}$ using its samples. In this section, some explicit formulas are given for approximations on this stationary distribution in which actual sampling on all $P_{k|k}$ is not required.

To investigate this approximation, the following results are at first established, which makes it clear that in finite recursions, the Homographic transformation of the robust state estimator RSEIO generally can not remove influences of its initial values.

\hspace*{-0.4cm}{\bf Lemma 3.} Assume that both $A^{[0]}$ and $A^{[1]}$ are invertible. Then, for arbitrary PDMs $X$ and $Y$ with a compatible dimension,
${\rm\bf H}_{m}\left(M^{[\star]},X\right)={\rm\bf H}_{m}\left(M^{[\star]},Y\right)$ if and only if $X=Y$, no matter $\star=1$ or $\star=0$.

\hspace*{-0.4cm}{\bf Proof:} From the definition of the Homographic transformation, direct algebraic manipulations show that when $X$ and $Y$ are positive definite and their dimensions are compatible,
\begin{eqnarray}
{\rm\bf H}_{m}\left(M^{[0]},X\right)-{\rm\bf H}_{m}\left(M^{[0]},Y\right)&=&\left[A^{[0]}XA^{[0]T}+G^{[0]}G^{[0]T}\right]-\left[A^{[0]}YA^{[0]T}+G^{[0]}G^{[0]T}\right]\nonumber\\
&=& A^{[0]}(X-Y)A^{[0]T} \\
{\rm\bf H}_{m}\left(M^{[1]},X\right)-{\rm\bf H}_{m}\left(M^{[1]},Y\right)&=&
\left[(A^{[1]}XA^{[1]T}+G^{[1]}G^{[1]T})^{-1}+H^{[1]T}H^{[1]}\right]^{-1}-\nonumber\\
& & \hspace*{1cm}
\left[(A^{[1]}YA^{[1]T}+G^{[1]}G^{[1]T})^{-1}+H^{[1]T}H^{[1]}\right]^{-1}\nonumber\\
& &\hspace*{-1cm}=
\left[I+(A^{[1]}XA^{[1]T}+G^{[1]}G^{[1]T})H^{[1]T}H^{[1]}\right]^{-1}A^{[1]}(X-Y)A^{[1]T}\times\nonumber\\
& & \hspace*{0.5cm}
\left[I+H^{[1]T}H^{[1]}(A^{[1]}YA^{[1]T}+G^{[1]}G^{[1]T})\right]^{-1}
\end{eqnarray}

The conclusions are immediate from these relations and the regularity of both $A^{[0]}$ and $A^{[1]}$. This completes the proof. \hspace{\fill}$\Diamond$

Assume that the Markov chain $\gamma_{k}$ is in its stationary state, and the PCM $P_{k|k}$ starts from $P^{\star}$. Let $P_{0|0},\;P_{1|1},\;\cdots P_{n|n}$ be its first $n+1$ samples, and consider all the possible values that these samples may take and the probability of their occurrence. Obviously from Lemma 3, when both $A^{[0]}$ and $A^{[1]}$ are invertible, there are $2^{k}$ possible values that $P_{k|k}$ may take, which is in accordance with all the realizations of the Markov chain $\gamma_{i}|_{i=1}^{k}$ with $\gamma_{i}\in\{0,\;1\}$. Recall that
${\rm\bf H}_{m}\left(M^{[1]},\;P^{\star}\right)=P^{\star}$. It is clear that for an arbitrary positive integer $k$,
\begin{equation}
{\rm\bf H}_{m}\left(M^{[1]k},\;P^{\star}\right)=
{\rm\bf H}_{m}\left[M^{[1](k-1)},{\rm\bf H}_{m}\left(M^{[1]},\;P^{\star}\right)\right]
={\rm\bf H}_{m}\left(M^{[1](k-1)},\;P^{\star}\right)
=\cdots=P^{\star}
\end{equation}
On the other hand, if it exists, let $P^{\S}$ denote the solution to the algebraic Lyapunov equation $P=A^{[0]}PA^{[0]T}+G^{[0]}G^{[0]T}$ that is positive definite. Then, from Lemma 3, it is clear that if $P\not\in\{P^{\star},\;P^{\S}\}$, then, ${\rm\bf H}_{m}\left(M^{[\#]},\;P\right)\neq P$, no matter $\#$ is equal to $1$ or $0$.

From these arguments, the following results can be obtained, while their proof is included in the appendix.

\hspace*{-0.4cm}{\bf Lemma 4:} Let ${\cal P}^{[n]}$ denote the set consisting of all possible values that $P_{k|k}|_{k=0}^{n}$ may take which has its initial value being $P^{\star}$ and recursively updates according to the stationary process of the Markov chain $\gamma_{k}$. Then, the number of the elements in ${\cal P}^{[n]}$ is equal to $2^{n}$ and the set ${\cal P}^{[n]}$ can be expressed as
\begin{equation}
{\cal P}^{[n]}=\left\{P^{\star},\;{\rm\bf H}_{m}\left(M^{[0]},\;P^{\star}\right)\right\}\bigcup\left\{\:P\:\left|\:\begin{array}{l}
P={\rm\bf H}_{m}\left[M^{[\gamma_{k}]}M^{[\gamma_{k-1}]}\cdots M^{[\gamma_{2}]},\;{\rm\bf H}_{m}\left(M^{[0]},\;P^{\star}\right)\right]\\
\hspace*{1cm}\gamma_{j}\in\{0,\;1\},\;j\in\{2,3,\cdots,k\},\;k\in\{2,3,\cdots,n\}\end{array}\:\right.\right\}
\end{equation}

For any sequence $\gamma_{j}|_{j=1}^{k}$ with $\gamma_{j}\in\{0,\;1\}$ and $k\in\{1,2,\cdots,n-1\}$, define $l(\gamma_{j}|_{j=1}^{k})$ and $\bar{P}^{[l(\gamma_{j}|_{j=1}^{k})]}$ respectively as
\begin{equation}
l(\gamma_{j}|_{j=1}^{k})=1+2^{k-1}+\sum_{j=1}^{k-1}\gamma_{j}2^{j-1},\hspace{0.5cm}
\bar{P}^{[l(\gamma_{j}|_{j=1}^{k})]}={\rm\bf H}_{m}\left(M^{[\gamma_{k}]}M^{[\gamma_{k-1}]}\cdots M^{[\gamma_{1}]}M^{[0]},\;P^{\star}\right)
\end{equation}
Moreover, define $\bar{P}^{[1]}=P^{\star}$. Then, from the proof of Lemma 4, it can be understood that ${\cal P}^{[n]}=\{\bar{P}^{[1]},\;\bar{P}^{[2]},\; \cdots,\;
\bar{P}^{[2^{n}]}\}$.

The following theorem gives a convergence value of $\frac{1}{n+1}\sum_{k=0}^{n}I_{{\cal B}_{\varepsilon}}(P_{k|k})$, which is helpful in deriving approximations for the stationary distribution of the PCM $P_{k|k}$. Its proof is given in the appendix.

\hspace*{-0.4cm}{\bf Theorem 3:} Let $\lceil\star\rceil$ denote the minimal integer that is not smaller than $\star$. For a prescribed positive $\varepsilon$, define set ${\cal N}_{\varepsilon}$ as
${\cal N}_{\varepsilon}=\left\{\:j\:\left|\:\bar{P}^{[j]}\in {\cal P}^{[n]}\bigcap {\cal B}_{\varepsilon}\right.\right\}$. Then,
\begin{equation}
\lim_{n\rightarrow\infty}\frac{1}{n+1}\sum_{k=0}^{n}I_{{\cal B}_{\varepsilon}}(P_{k|k})
=\lim_{n\rightarrow\infty}\sum_{j\in{\cal N}_{\varepsilon}}\left(1-\gamma_{st}^{n-\lceil log_{2}(j) \rceil}\right)\gamma_{st}^{\sum_{i=1}^{\lceil log_{2}(j) \rceil}\gamma_{i}^{[j]}}(1-\gamma_{st})^{\lceil log_{2}(j) \rceil-\sum_{i=1}^{\lceil log_{2}(j) \rceil}\gamma_{i}^{[j]}}
\label{eqn:8}
\end{equation}
in which $\gamma_{st}$ stands for the probability that $\gamma_{k}$ takes the value of $1$ at its stationary state, and $\gamma_{i}^{[j]}$ is the binary code for
$j-1-2^{\lceil log_{2}(j) \rceil-1}$.

In the above theorem, an explicit formula is given for the stationary distribution of the PCM of the robust state estimator RSEIO. In principle, its value can be computed for each prescribed $\varepsilon$, which means that this distribution function can be obtained to an arbitrary accuracy, provided that a computer with a sufficient computation speed and a sufficient memory capacity is available. Note that the value of $2^{n}$ increases exponentially with the increment of the sample size $n$ and a large $n$ is generally appreciated as it leads to a more accurate approximation on the distribution function of the stationary PCM. It appears reasonable to claim that in general, conclusions of the above theorem can not be directly utilized in actual computations, and some other more efficient approximations are still required.

From Equation (\ref{eqn:8}), however, it is obvious that when $\gamma_{st}$ is approximately equal to $1$, $(1-\gamma_{st})^{\lceil log_{2}(j) \rceil}$ is very small if the corresponding $\gamma_{i}^{[j]}|_{i=1}^{\lceil log_{2}(j) \rceil}$ has many zeros. On the other hand, from the proof of Theorem 1, it is clear that when the length of the sequence $\gamma_{i}^{[j]}|_{i=1}^{\lceil log_{2}(j) \rceil}$, that is, $\lceil log_{2}(j) \rceil$, is large, the probability that it has a large number of zeros is high. These mean that contributions of an element $P^{j}\in{\cal P}^{[n]}$ with a large $j$ to the stationary distribution of the random PCM process are usually very small and can therefore be neglected. On the basis of these observations, the following approximation is developed for this stationary distribution which is given in Theorem 4.

Note that ${\rm\bf H}_{m}\left(M^{[0]},X\right)=A^{[0]}XA^{[0]T}+G^{[0]}G^{[0]T}$. It is straightforward to prove from the definition of the Riemannian distance that for an arbitrary PDM $X$, there exist finite positive numbers $a$ and $b$ that do not depend on the matrix $X$, such that
\begin{equation}
\delta\left[{\rm\bf H}_{m}\left(M^{[0]},X\right),\;P^{\star}\right]\leq a\delta(X,P^{\star})+b
\label{eqn:3}
\end{equation}

On the other hand, let $N_{0}^{[m]}(j)$ denote the number of zeros of a particular finite length binary sequence $\gamma_{i}^{[j]}|_{i=1}^{m}$ with $\gamma_{i}^{[j]}\in\{0,\;1\}$ and $m\leq n$, in which $n$ stands for the PCM sample length. Then, when the Markov chain $r_{k}|_{k=1}^{\infty}$ is in its stationary state, the occurrence probability of this sequence is equal to $\gamma_{st}^{m-N_{0}^{[m]}(j)}(1-\gamma_{st})^{N_{0}^{[m]}(j)}$ in which $\gamma_{st}$ is the stationary probability for $\gamma_{k}=1$. When $\gamma_{st}$ is approximately equal to $1$, this number dramatically decreases with the increment of $N_{0}^{[m]}(j)$. Assume that a PCM with probability smaller than $\varepsilon_{p}$ can be neglected without significant influences on the stationary distribution of the random process $P_{k|k}$. Then, from $\gamma_{st}^{m-N_{0}^{[m]}(j)}(1-\gamma_{st})^{N_{0}^{[m]}(j)}\leq \varepsilon_{p}$, it can be directly proved that in all the binary sequences of length $m$, only these with
\begin{equation}
N_{0}^{[m]}(j)\leq\frac{ln(\varepsilon_{p})-k\:ln(\gamma_{st})}{ln(1-\gamma_{st})-ln(\gamma_{st})}
\end{equation}
lead to a PCM that should be considered in establishing the stationary distribution of the random process $P_{k|k}$.

In addition, for a binary sequence of length $k$, say, $\gamma_{i}|_{i=1}^{k}$, assume that $\gamma_{t_{j}}=0$, $j=1,2,\cdots,p$, with $1\leq t_{1}< t_{2}<\cdots< t_{p}\leq k$. Consider the distance between the corresponding PCM $P_{k|k}$ and the matrix $P^{\star}$. Note that
\begin{equation}
\delta\left[{\rm\bf H}_{m}\left(M^{[1]},X\right),\;P^{\star}\right]=\delta\left[{\rm\bf H}_{m}\left(M^{[1]},X\right),\;{\rm\bf H}_{m}\left(M^{[1]},P^{\star}\right)\right]\leq\alpha_{1h}\delta(X,P^{\star})
\end{equation}
is valid for an arbitrary PDM $X$. A repetitive utilization of this relation leads to that for any positive integer $m$ and any PDM $X$,
\begin{eqnarray}
\delta\left[{\rm\bf H}_{m}\left(M^{[1]m},X\right),\;P^{\star}\right]&=&\delta\left\{{\rm\bf H}_{m}\left[M^{[1]},\;{\rm\bf H}_{m}\left(M^{[1](m-1)},X\right)\right],\;{\rm\bf H}_{m}\left(M^{[1]},P^{\star}\right)\right]\nonumber\\
&\leq&\alpha_{1h}\delta\!\left\{{\rm\bf H}_{m}\left(M^{[1](m-1)},X\right),\;P^{\star}\right\}\nonumber\\
&=&\cdots \nonumber\\
&=& \alpha_{1h}^{m}\delta(X,P^{\star})
\end{eqnarray}

From this inequality and Equation (\ref{eqn:3}), the following inequality is obtained
\begin{eqnarray}
& & \delta\left[{\rm\bf H}_{m}\!\left(\prod_{i=k}^{1}\!\!M^{[\gamma_{i}]},X\right),\;P^{\star}\right]\nonumber\\
&=&
\delta\!\left[{\rm\bf H}_{m}\!\left(\prod_{i=k}^{t_{p}+1}\!\!M^{[\gamma_{i}]}\!\times\! M^{[0]}\!\times \!\prod_{i=t_{p}-1}^{t_{p-1}+1}\!\!M^{[\gamma_{i}]}\!\times M^{[0]}\!\times \cdots \!\prod_{i=t_{2}-1}^{t_{1}+1}\!\!M^{[\gamma_{i}]}\!\times M^{[0]}\!\!\times\!\! \prod_{i=t_{1}-1}^{1}M^{[\gamma_{i}]},X\right),\;P^{\star}\right]\nonumber\\
&=&\delta\!\left\{{\rm\bf H}_{m}\left[M^{[1](k-t_{p})},\;{\rm\bf H}_{m}\left(\prod_{j=p}^{1}\left(M^{[0]}M^{[1](t_{j}-t_{j-1}-1)}\right),\;X\right)\right],\;P^{\star}\right\}\nonumber\\
&\leq &\alpha_{1h}^{k-t_{p}}\delta\!\left[{\rm\bf H}_{m}\!\left(\prod_{j=p}^{1}\left(M^{[0]}M^{[1](t_{j}-t_{j-1}-1)}\right),\;X\right),\;P^{\star}\right]\nonumber\\
&=& \alpha_{1h}^{k-t_{p}}\delta\!\left\{{\rm\bf H}_{m}\!\left[M^{[0]},\;{\rm\bf H}_{m}\left(\prod_{j=p}^{2}\left(M^{[1](t_{j}-t_{j-1}-1)}M^{[0]}\right)M^{[1](t_{1}-1)},\;X\right)\right],\;P^{\star}\right\}\nonumber\\
&\leq & \alpha_{1h}^{k-t_{p}}\left\{a\delta\!\left[{\rm\bf H}_{m}\!\left(\prod_{j=p}^{2}\left(M^{[1](t_{p}-t_{p-1}-1)}M^{[0]}\right)M^{[1](t_{1}-1)},\;X\right),\;P^{\star}\right]+b\right\}\nonumber\\
&=&\alpha_{1h}^{k-t_{p}}b+\alpha_{1h}^{k-t_{p}}a\delta\!\left\{{\rm\bf H}_{m}\!\left[M^{[1](t_{p}-t_{p-1}-1)},\; {\rm\bf H}_{m}\!\left(\prod_{j=p-1}^{1}\left(M^{[0]}M^{[1](t_{j-1}-t_{j-2}-1)}\right),X\right)\right],\;P^{\star}\right\}\nonumber\\
&\leq & \cdots \nonumber\\
&\leq& \sum_{j=0}^{p}\alpha_{1h}^{k-t_{p-j}}a^{j}b+\alpha_{1h}^{k}a^{p}\delta(X,P^{\star})
\label{eqn:4}
\end{eqnarray}
in which $t_{0}$ is defined as $t_{0}=0$. Hence, if $k>>t_{p}$, then
\begin{equation}
\delta\!\left[{\rm\bf H}_{m}\left(\prod_{i=k}^{1}M^{[\gamma_{i}]},X\right),\;P^{\star}\right]\approx 0
\end{equation}
This means that the PCM $P_{k|k}$ has a distinguishable distance from $P^{\star}$ only if $t_{p}\approx k$. Moreover, if $t_{l}\approx t_{l+1}\approx\cdots t_{m}\approx k$ and $t_{l-1}>>1$, it can be proved through similar arguments that
\begin{equation}
\delta\!\left[{\rm\bf H}_{m}\left(\prod_{i=k}^{1}M^{[\gamma_{i}]},X\right),\;{\rm\bf H}_{m}\left(M^{[0](k-l)},\; P^{\star}\right)\right]\approx 0
\end{equation}
That is, the PCM $P_{k|k}$ is approximately equal to ${\rm\bf H}_{m}\left(M^{[0](k-l)},\; P^{\star}\right)$.

Recall that ${\rm\bf H}_{m}\left(M^{[1]},\; P^{\star}\right)=P^{\star}$ and ${\rm\bf H}_{m}\left(M^{[0]},\; P^{\star}\right)=P^{\S}\neq P^{\star}$. The above arguments and Theorem 3 suggest that when the random process $P_{k|k}$ is initialized with $P^{\star}$, then, after the first occurrence of $\gamma_{k}=0$, the succeeding $P_{k|k}$ intends to converges to one of the elements of the set ${\cal P}^{[\infty]}$ which is defined as ${\cal P}^{[\infty]}=\{P^{\star}\}\bigcup
\left\{\:P\:\left|\:P={\rm\bf H}_{m}\left(M^{[0]i},\; P^{\star}\right),\;i=1,2,\cdots\:\right.\right\}$. In other words, with a high probability, the random matrix $P_{k|k}$ is concentrated around the elements of the set ${\cal P}^{[\infty]}$, and a PCM far away from every element of this set usually has a negligible probability to occur. These concentrations become more dominating if both $a$ and $b$ are not very large and $\alpha_{1h}$ is significantly smaller than $1$, which can be understood from Equation (\ref{eqn:4}).

From these observations, it seems reasonable to approximate the support of $P_{\infty|\infty}$ by the set ${\cal P}^{[\infty]}$. When this approximation is valid, a very simple and explicit formula can be derived for the stationary distribution of the random PCM process, which is given in the next theorem. Its proof is deferred to the appendix.

\hspace*{-0.4cm}{\bf Theorem 4:} Assume that the set ${\cal P}^{[\infty]}$ is a good approximation for the support of the stationary process of the PCM of the robust state estimator RSEIO. Then,
\begin{equation}
{\rm\bf P}_{r}(P_{\infty|\infty}=P^{\star})\approx \gamma_{st},\hspace{0.5cm}
{\rm\bf P}_{r}\left\{P_{\infty|\infty}={\rm\bf H}_{m}\left(M^{[0]i},\; P^{\star}\right)\right\}\approx \gamma_{st}(1-\gamma_{st})^{i},\;\;i=1,2,\cdots
\end{equation}

Note that when $\gamma_{st}\approx 1$, $\gamma_{st}(1-\gamma_{st})^{i}$ decreases rapidly to $0$ with the increment of the index $i$. This means that when the data arrival probability in the stationary PCM process is high, only a few elements of the set ${\cal P}^{[\infty]}$, that is, ${\rm\bf H}_{m}\left(M^{[0]i},\; P^{\star}\right)$,  are required in computing the approximation for the stationary distribution of the random PCM process. Another attractive characteristic of this approximation is that its accuracy does not depend on the length of time series PCM samples, and therefore can greatly reduce computation burdens.

While Theorem 4 provides a very simple approximation, it is still a challenging problem to derive its approximation accuracy, as well as explicit conditions on system parameters under which the delta function approximation is valid.

\section{A Numerical Example}

To illustrate accuracies of the derived approximations, various numerical simulations have been performed. Some typical results are reported in this
section. The adopted plant is a modification of that utilized in \cite{zhou14}
which has the following system matrices, initial conditions, and
covariance matrices for process noises and measurement errors,
respectively.
\begin{eqnarray*}
& & A_{k}(\varepsilon_{k})\!=\!\left[\!\!\begin{array}{cc} 1.1234 & 0.0196 \\
0 & 0.9802 \end{array}\!\!\right]\!+\!\left[\!\!\begin{array}{c} 0.0198 \\
0
\end{array}\!\!\right]\!\varepsilon_{k}\left[0\;\; 5\right],\hspace{0.2cm}
B_{k}(\varepsilon_{k})\!=\!\left[\!\!\begin{array}{cc} 1 & 0 \\ 0 &
1
\end{array}\!\!\right],\hspace{0.2cm} Q_{k}\!=\!\left[\!\!\begin{array}{cc}
1.9608 & 0.0195 \\ 0.0195 & 1.9605 \end{array}\!\!\right] \\
& & C_{k}(\varepsilon_{k})=[1\;\; -1],\hspace{0.5cm}
R_{k}=1,\hspace{0.5cm} {\rm\bf
E}\{x_{0}\}=[1\;\;0]^{T},\hspace{0.5cm} P_{0}=I_{2}
\end{eqnarray*}
in which $\varepsilon_{k}$ stands for a time varying parametric
error that is independent of each other and has a uniform
distribution over the interval $[-1,\;1]$. The measurement
dropping process $\gamma_{k}$ is assumed to be a Markov chain. Moreover, the estimator design
parameter $\mu_{k}$ is selected as $\mu_{k}\equiv 0.8$.

The only modification is made on the matrix $A_{k}(\varepsilon_{k})$ which makes the nominal system unstable. This makes the simulation system more appropriate in  investigating typical behaviors of a state estimator with random data loss, noting that if the nominal system is stable, the PCM of the robust state estimator RSEIO converges to a constant PDM with the increment of the temporal variable $k$, even all the measured data are lost \cite{ms12,sem10,ssfpjs04,zhou14}.

Direct numerical computations show that for this system, both $A^{[0]}$ and $A^{[1]}$ are invertible, and $(A^{[1]}, G^{[1]})$ is controllable, while $(H^{[1]}, A^{[1]})$ is observable. Moreover, using the Matlab command {\it dric.m}, the following $P^{\star}$ is obtained,
\begin{displaymath}
P^{\star}=\left[\begin{array}{cc}21.3283 & 20.2784 \\ 20.2784 & 20.0754\end{array}\right]
\end{displaymath}

Various situations have been tested on this numerical example. The obtained computation results confirm the theoretical results established in the previous sections. In these simulations, both empirical stationary distribution of the random PCM process and its approximations based respectively on the ergodicity property of this random process and the delta functions are computed. In computing the empirical stationary distribution, $5\times 10^{4}$ trials are performed for simulating the PCM at $k=10^{3}$ that are initialized with ${\rm\bf P}_{r}(\gamma_{0}=1)=0.7$ and a PDM $P_{0|0}=10^{3}I_{2}$, and the empirical stationary distribution is calculated using the obtained $\delta(P_{10^{3}|10^{3}},\;P^{\star})$. When the approximation of Theorem 2 is used, the PCM $P_{k|k}$ is initialized with $P_{0|0}=P^{\star}$ and
${\rm\bf P}_{r}(\gamma_{0}=1)=\frac{1-\alpha}{2-\alpha-\beta}$ which is the stationary distribution of the Markov chain $\gamma_{k}$. The first $5\times 10^{4}$ samples of the PCM, that is, $P_{k|k}|_{k=1}^{5\times 10^{4}}$, are simulated which are further used to compute an approximation of the stationary distribution of the PCM on the basis of Theorem 2.

In computing the empirical stationary distribution and its Theorem 2 based approximation, an interval $[0,\;\delta_{max}]$ is at first divided into $N_{e}$ intervals of an equal length, in which $N_{e}$ and $\delta_{max}$ are respectively a prescribed positive integer and a prescribed positive number that are suitably selected according to the maximal value of the distance from the simulated PCMs to the matrix $P^{\star}$. Then, the number of the simulated PCM samples are counted that satisfy $i\frac{\delta_{max}}{N_{e}}\leq \delta(P_{10^{3}|10^{3}},\;P^{\star})<(i+1)\frac{\delta_{max}}{N_{e}}$ for the empirical stationary distribution, and
$i\frac{\delta_{max}}{N_{e}}\leq \delta(P_{k|k},\;P^{\star})<(i+1)\frac{\delta_{max}}{N_{e}}$ for the Theorem 2 based approximation, $i=0,1,\cdots,N_{e}-1$. Finally, this number is divided by the total number of the simulated samples, that is, $5\times 10^{4}$, and is regarded to be a value proportional to that of the empirical probability density function (PDF) of the stationary PCM process and its Theorem 2 based approximation at $\delta=\left(i+\frac{1}{2}\right)\frac{\delta_{max}}{N_{e}}$, and the corresponding points are connected using the Matlab command {\it plot.m}. The obtained curves are regarded to be proportional to those of the empirical PDF and its approximation using Theorem 2 (The proportional rate is $\frac{\delta_{max}}{N_{e}}$.). To make statements concise, with a little abuse of terminology, these curves are respectively called empirical PDF and its approximation.

When the approximation of Theorem 4 is used, the following method is adopted for comparing results of empirical distributions of the stationary PCM process and its approximations. At first, select a suitable positive integer $N_{d}$, and compute $d_{i}=\delta({\rm\bf H}_{m}(M^{[0]i},\;P^{\star}),\;P^{\star})$ and  $p_{i}=\gamma_{st}(1-\gamma_{st})^{i}$, $i=1,2,\cdots,N_{d}$. This $N_{d}$ is chosen to guarantee that $\gamma_{st}(1-\gamma_{st})^{N_{d}+1}$ is smaller than some threshold values, for example, $10^{-7}$. Then, another positive integer $N_{s}$ is selected according to the distance distribution between the simulated PCM and the matrix $P^{\star}$, which is used to reflect the closeness of the simulated stationary distribution to delta functions. Afterwards, the number is counted of the simulated PCMs that has a distance to the matrix $P^{\star}$ belonging to the interval ${\cal I}_{i}$, $i=0,1,\cdots,N_{d}$, in which
\begin{displaymath}
{\cal I}_{i}=\left\{\begin{array}{lll} \left[0,\; \frac{d_{1}}{N_{s}}\right], &\;\;\; &i=0\\
\left(d_{i}-\frac{d_{i}-d_{i-1}}{N_{s}},\; d_{i}+\frac{d_{i+1}-d_{i}}{N_{s}}\right], &\;\;\; &i=1,2,\cdots,N_{d}-1 \\
\left(d_{N_{d}}-\frac{d_{N_{d}}-d_{N_{d}-1}}{N_{s}},\; d_{N_{d}}\right], &\;\;\; &i=N_{d}  \end{array}\right.
\end{displaymath}
Finally, these numbers are divided by the total number of the simulated samples, that is, $5\times 10^{4}$, and regarded to be the empirical value of the probability of the stationary PCM process and its Theorem 2 based approximation at $\delta(P_{\infty|\infty},\;P^{\star})=d_{i}$, $i=0,1,\cdots,N_{d}$. Clearly, under the condition that these values are close to $p_{i}$, the greater the integer $N_{s}$ is, the closer the stationary distribution of the random PCM process to delta functions.

\begin{figure}[!ht]
\begin{center}
\hspace*{-0.4cm}\includegraphics[width=3.0in]{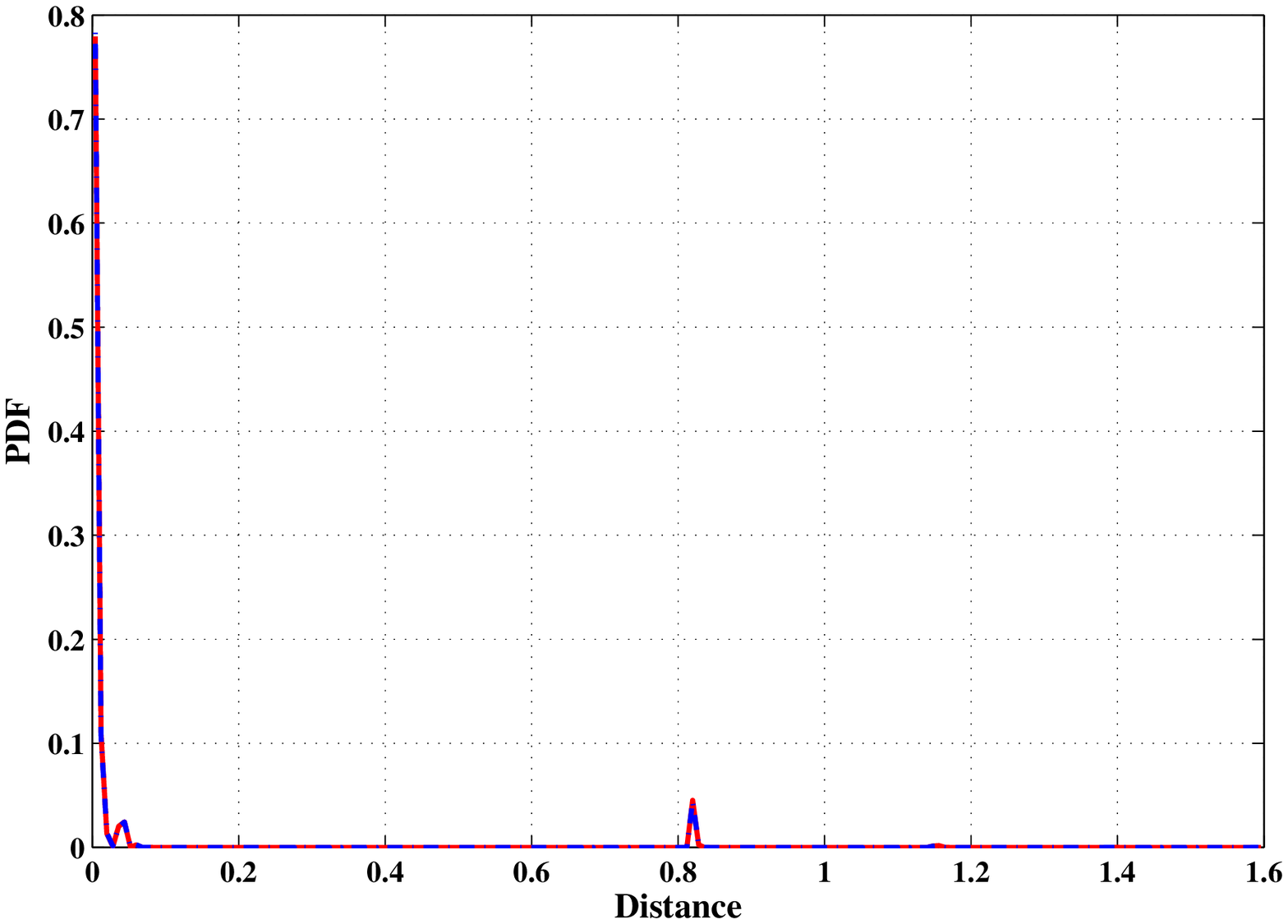}
\hspace{0.4cm}\includegraphics[width=3.0in]{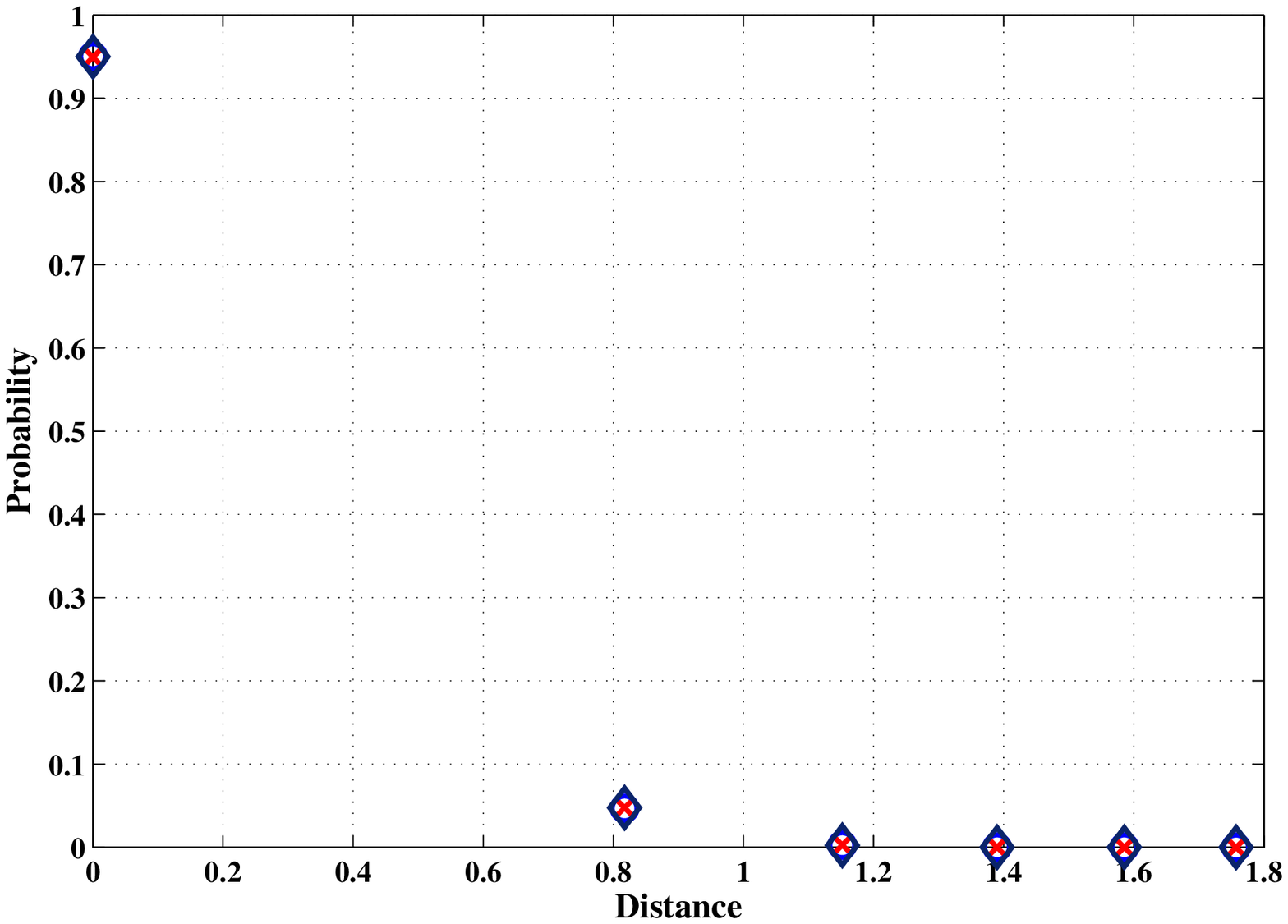}

\vspace{-0.2cm} \hspace*{-1.0cm} {\footnotesize (a)
probability density function} \hspace{5.0cm}
{\footnotesize (b) probability}
\end{center}

\vspace{-0.5cm} \caption{Empirical PDF and probability of the stationary PCM process, together with their approximations. ($\alpha=0.95$, $\beta=0.05$) $-\!\!\!-\!\!\!-\!\!\!-$: empirical PDF; $-\:\cdot\:-$: PDF approximation based on Theorem 2; ${\rm X}$: empirical probability; $\circ$: probability approximation based on Theorem 2; $\Diamond$: probability approximation based on Theorem 4.}
\end{figure}

In Figure 1a, simulations results with $\alpha=0.95$ and $\beta=0.05$ are plotted for the empirical PDF of the stationary PCM process and its Theorem 2 based approximation, in which $\delta_{max}$ and $N_{e}$ are respectively chosen as $\delta_{max}=1.6$ and $N_{e}=200$. The corresponding empirical probability is plotted in Figure 1b, together with its approximations based respectively on Theorems 2 and 4, in which $N_{d}$ and $N_{s}$ respectively take the value $N_{d}=5$ and $N_{s}=10$. To understand the approximation accuracy more clearly, the computed values used in plotting Figure 1b are given in Table I.

\begin{table}[!ht]
\caption{Empirical Probability and Its Approximations ($\alpha=0.95$, $\beta=0.05$) }
\vspace{-0.5cm}
\begin{center}
\begin{tabular}{|p{2.5cm}|p{2.5cm}|p{2.5cm}|p{2.5cm}|}
\hline
Distance ($d_{i}$) & App.Theorem 4 & App.Theorem 2 & Empirical Prob. \\
\hline
\hline
  $0$  & $9.5000\times 10^{-1}$ & $9.5044\times 10^{-1}$ & $9.4940\times 10^{-1}$ \\
  \hline
  $8.1725\times 10^{-1}$ & $4.7500\times 10^{-2}$ & $4.6760\times 10^{-2}$ & $4.7560\times 10^{-2}$ \\
  \hline
  $1.1519$  & $2.3750\times 10^{-3}$ & $2.6400\times 10^{-3}$ &  $2.9200\times 10^{-3}$ \\
  \hline
  $1.3900$ & $1.1875\times 10^{-4}$ & $1.4000\times 10^{-4}$ & $1.2000\times 10^{-4}$ \\
  \hline
  $1.5855$ & $5.9375\times 10^{-6}$ & $2.0000\times 10^{-5}$ & $0$ \\
  \hline
  $1.7572$ & $2.9688\times 10^{-7}$ & $0$     &  $0$ \\
\hline
\end{tabular}
\end{center}
\end{table}

From these results, it is clear that when the data loss probability is low in the stationary state of the Markov chain $\gamma_{k}$, which corresponds to a large $\alpha$ and a small $\beta$, the PDF of the stationary PCM process is really very close to a series of delta functions, and the approximation based on either Theorem 2 or Theorem 4 has a high accuracy.

\begin{figure}[!ht]
\begin{center}
\hspace*{-0.4cm}\includegraphics[width=3.0in]{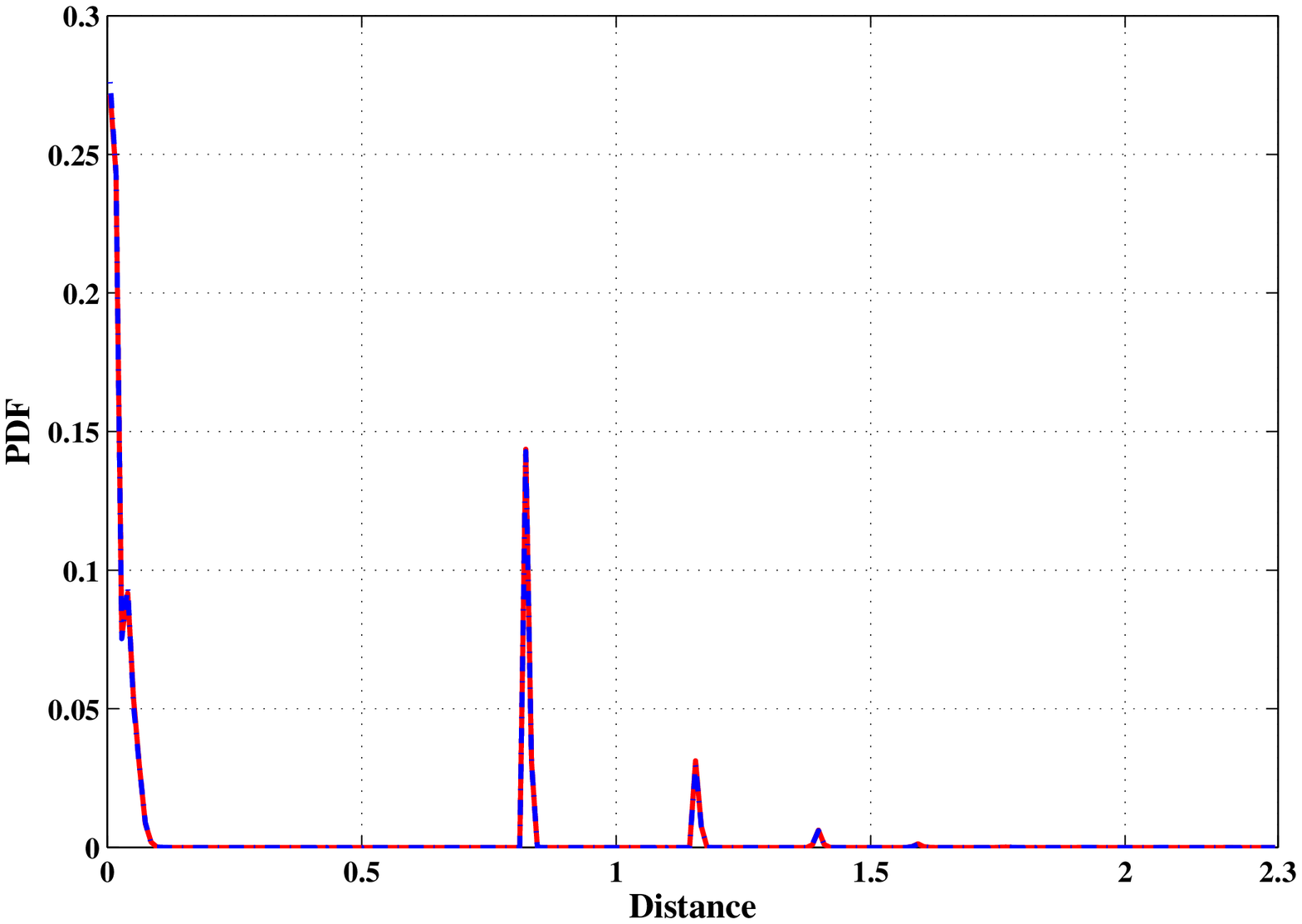}
\hspace{0.4cm}\includegraphics[width=3.0in]{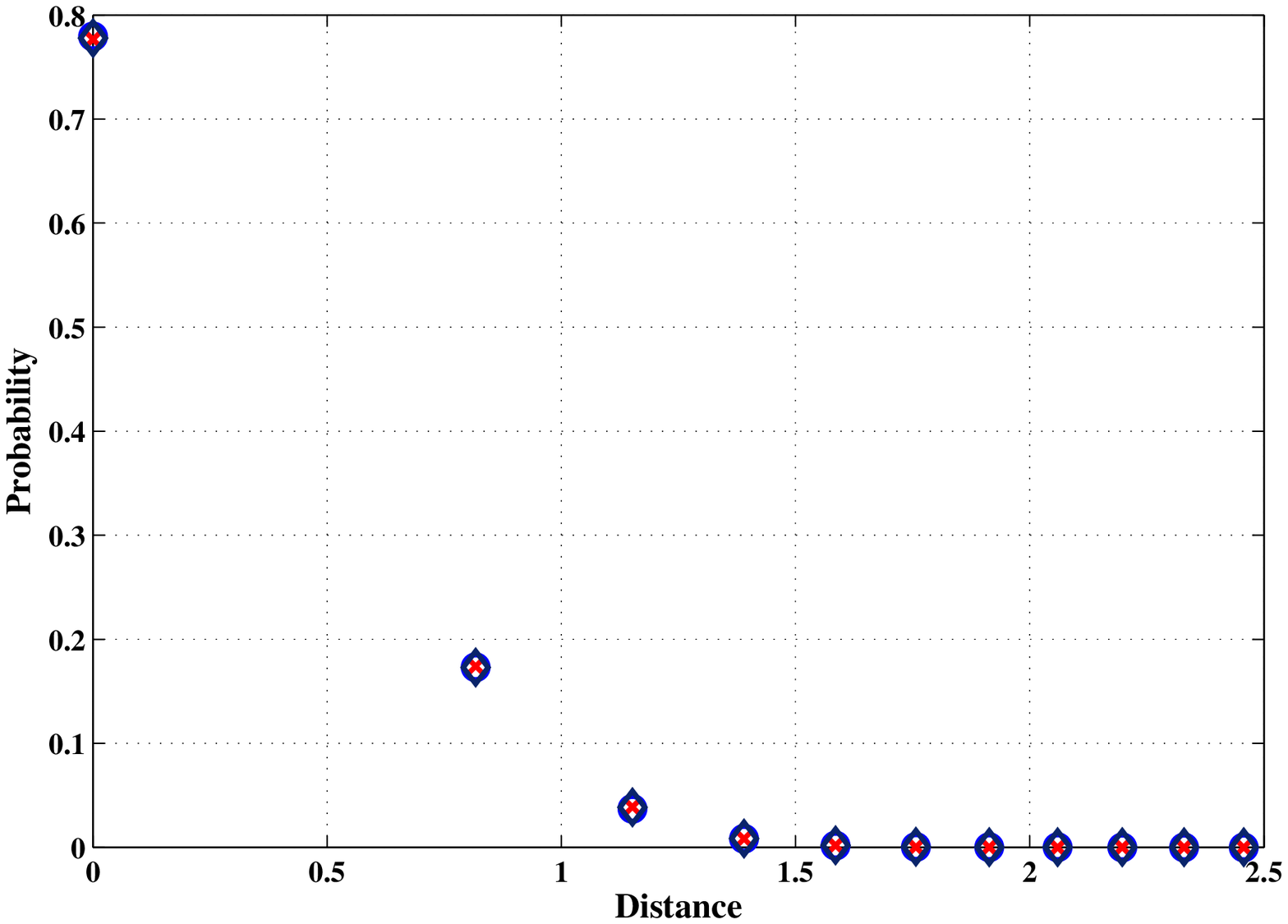}

\vspace{-0.2cm} \hspace*{-1.0cm} {\footnotesize (a)
probability density function} \hspace{5.0cm}
{\footnotesize (b) probability}
\end{center}

\vspace{-0.5cm} \caption{Empirical PDF and probability of the stationary PCM process, together with their approximations. ($\alpha=0.80$, $\beta=0.30$) $-\!\!\!-\!\!\!-\!\!\!-$: empirical PDF; $-\:\cdot\:-$: PDF approximation based on Theorem 2; ${\rm X}$: empirical probability; $\circ$: probability approximation based on Theorem 2; $\Diamond$: probability approximation based on Theorem 4.}
\end{figure}

When $\alpha=0.80$ and $\beta=0.30$, the corresponding simulated results are given in Figure 2 and Table II. In the related computations, $\delta_{max}=2.3$,  $N_{e}=200$, $N_{d}=10$ and $N_{s}=9$ are utilized. These results show that when the data loss probability has a moderate value, approximation of the stationary distribution of the random PCM process by delta functions is still of a high accuracy.

\begin{table}[!ht]
\caption{Empirical Probability and Its Approximations ($\alpha=0.80$, $\beta=0.30$)}
\begin{center}
\vspace{-0.5cm}
\begin{tabular}{|p{2.5cm}|p{2.5cm}|p{2.5cm}|p{2.5cm}|}
\hline
Distance ($d_{i}$) & App.Theorem 4 & App.Theorem 2 & Empirical Prob. \\
\hline
\hline
          $0$        &     $7.7778\times 10^{-1}$ & $7.7920\times 10^{-1}$ & $7.7648\times 10^{-1}$ \\ \hline
  $8.1725\times 10^{-1}$&  $1.7284\times 10^{-1}$ & $1.7318\times 10^{-1}$ & $1.7416\times 10^{-1}$ \\ \hline
  $1.1519$ &               $3.8409\times 10^{-2}$ & $3.7120\times 10^{-2}$ & $3.8780\times 10^{-2}$ \\ \hline
  $1.3900$ &               $8.5353\times 10^{-3}$ & $8.2800\times 10^{-3}$ & $8.1200\times 10^{-3}$ \\ \hline
  $1.5855$ &               $1.8967\times 10^{-3}$ & $1.5600\times 10^{-3}$ & $1.8000\times 10^{-3}$ \\ \hline
  $1.7572$ &               $4.2150\times 10^{-4}$ & $3.0000\times 10^{-4}$ & $2.2000\times 10^{-4}$ \\ \hline
  $1.9136$ &               $9.3666\times 10^{-5}$ & $1.0000\times 10^{-4}$ & $6.0000\times 10^{-5}$ \\ \hline
  $2.0595$ &               $2.0815\times 10^{-5}$ & $2.0000\times 10^{-5}$ & $            0       $ \\ \hline
  $2.1975$ &               $4.6255\times 10^{-6}$ & $    0 $               & $2.0000\times 10^{-5}$ \\ \hline
  $2.3296$ &               $1.0279\times 10^{-6}$ & $    0 $               &           $0$    \\ \hline
  $2.4570$ &               $2.2842\times 10^{-7}$ & $    0 $               &           $0$ \\
\hline
\end{tabular}
\end{center}
\end{table}

Our experiences show that even when the measured data has a high probability to be lost, which corresponds to a small $\alpha$ and a large $\beta$, the approximation of Theorem 4 still has a good accuracy. Figure 3 and Table III give some simulation results with $\alpha=0.08$ and $\beta=0.92$, in which $\delta_{max}=20$,  $N_{e}=2000$, $N_{d}=40$ and $N_{s}=2$ are utilized. Clearly, the approximation of Theorem 2 still has a value close to the empirical distributions of the stationary PCM process, but relative errors of the approximation based on Theorem 4 becomes greater, especially when the distance $d_{i}$ is large. This can be seen from Figure 3a, which indicates  that when the distance is large, the empirical PDF is no longer a series of delta functions. In addition, even when the distance is near $0$, separations among successive delta functions become short, and the width of each delta function increases. All these factors affect approximation accuracy of Theorem 4. Despite these influences, it appears that the approximation accuracy is still acceptable.

\begin{figure}[!ht]
\begin{center}
\hspace*{-0.4cm}\includegraphics[width=3.0in]{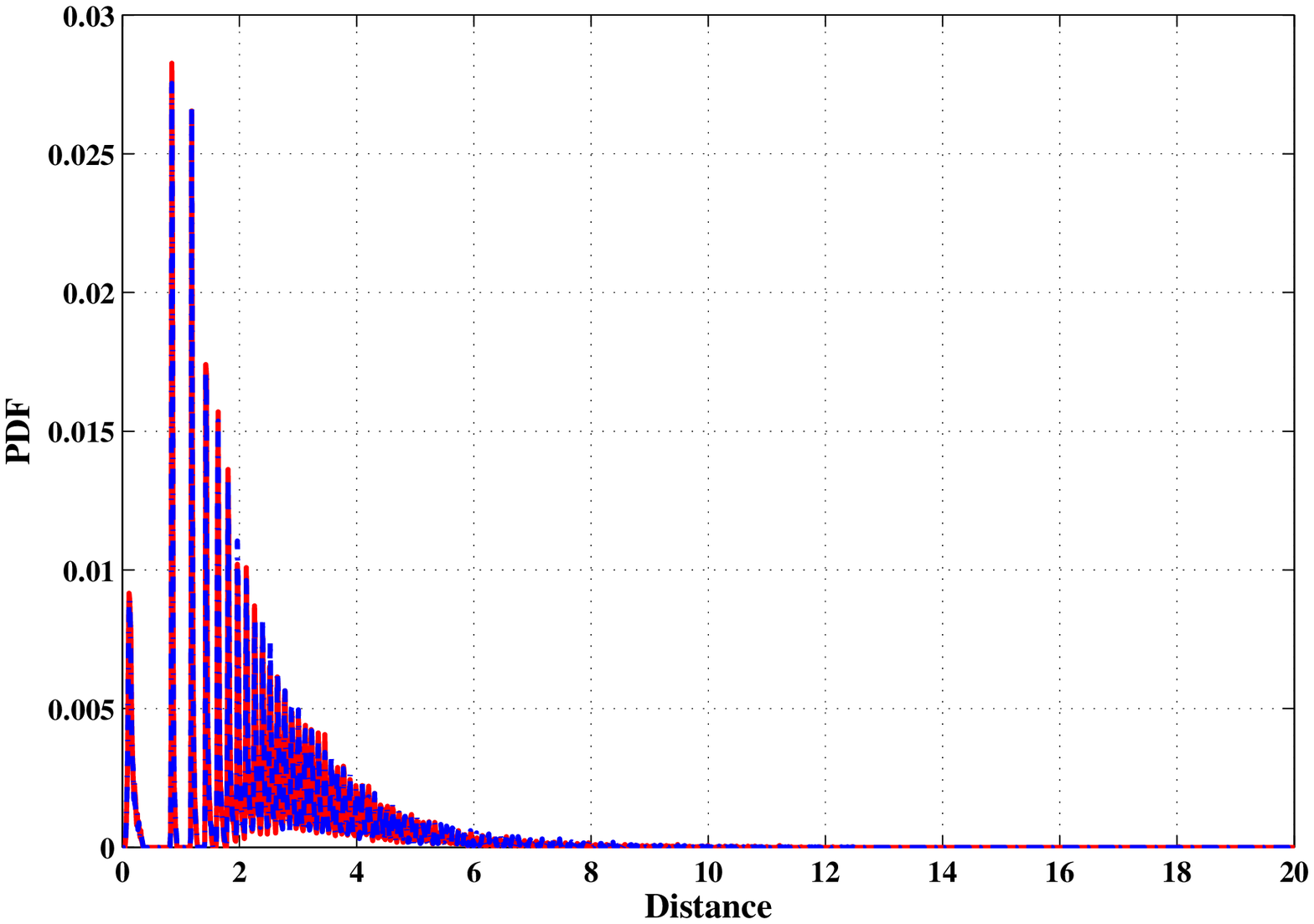}
\hspace{0.4cm}\includegraphics[width=3.0in]{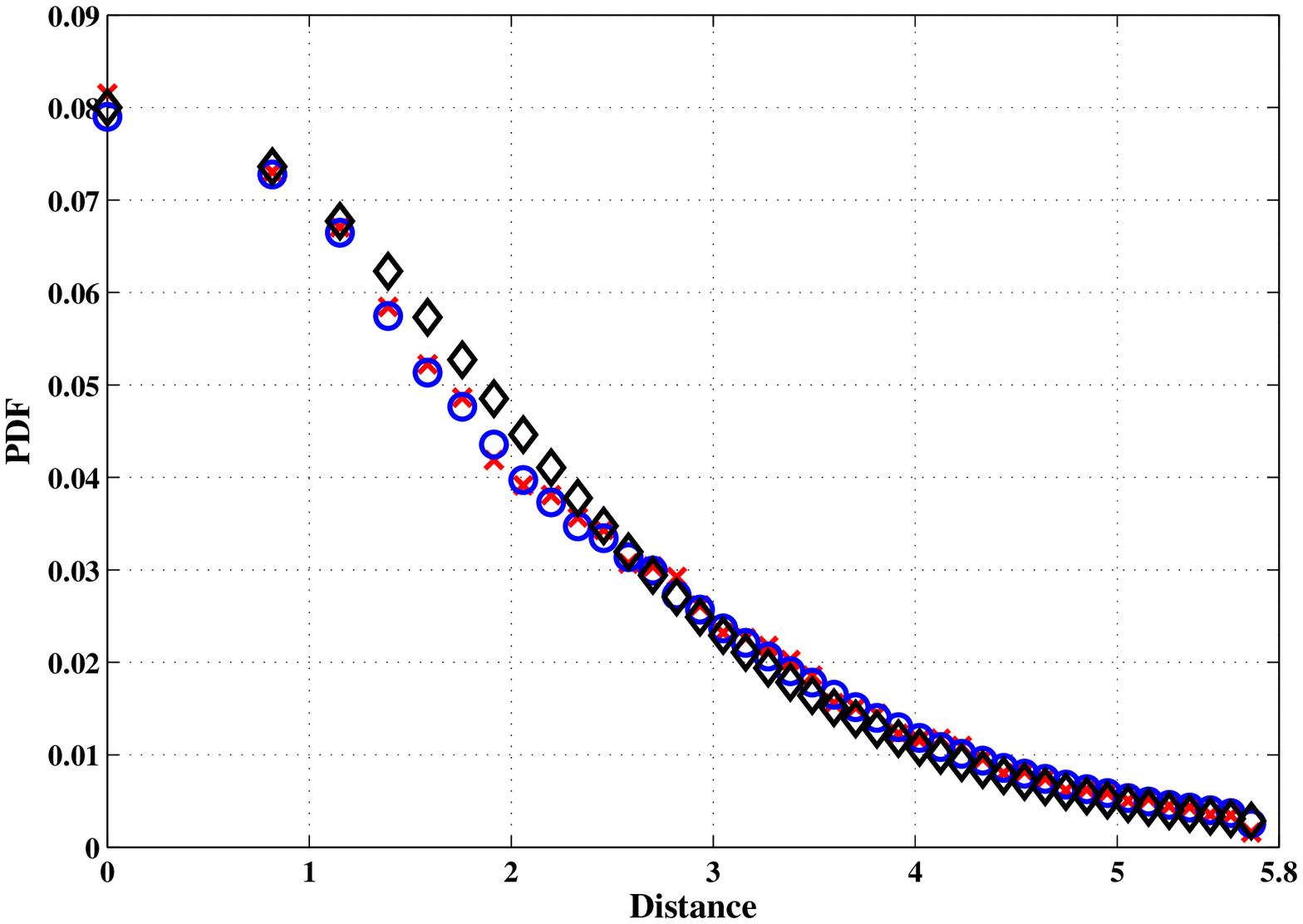}

\vspace{-0.2cm} \hspace*{-1.0cm} {\footnotesize (a)
probability density function} \hspace{5.0cm}
{\footnotesize (b) probability}
\end{center}

\vspace{-0.5cm} \caption{Empirical PDF and probability of the stationary PCM process, together with their approximations. ($\alpha=0.08$, $\beta=0.92$) $-\!\!\!-\!\!\!-\!\!\!-$: empirical PDF; $-\:\cdot\:-$: PDF approximation based on Theorem 2; ${\rm X}$: empirical probability; $\circ$: probability approximation based on Theorem 2; $\Diamond$: probability approximation based on Theorem 4.}
\end{figure}

\begin{table}[!ht]
\caption{Empirical Probability and Its Approximations ($\alpha=0.08$, $\beta=0.92$)}
\begin{center}
\vspace{-0.5cm}
\begin{tabular}{|p{2.5cm}|p{2.5cm}|p{2.5cm}|p{2.5cm}|}
\hline
Distance ($d_{i}$) & App.Theorem 4 & App.Theorem 2 & Empirical Prob. \\
\hline
\hline
  $0$ & $8.0000\times 10^{-2}$   &  $7.9000\times 10^{-2}$   &  $8.1500\times 10^{-2}$ \\ \hline
  $8.1725\times 10^{-1}$  & $7.3600\times 10^{-2}$  & $7.2760\times 10^{-2}$  & $7.3020\times 10^{-2}$ \\ \hline
  $1.1519$  & $6.7712\times 10^{-2}$  & $6.6460\times 10^{-2}$  & $6.7040\times 10^{-2}$ \\ \hline
  $1.3900$  & $6.2295\times 10^{-2}$  & $5.7460\times 10^{-2}$  & $5.8440\times 10^{-2}$ \\   \hline
  $1.5855$  & $5.7311\times 10^{-2}$  & $5.1340\times 10^{-2}$  & $5.2220\times 10^{-2}$ \\ \hline
  $1.7572$  & $5.2727\times 10^{-2}$  & $4.7640\times 10^{-2}$  & $4.8640\times 10^{-2}$ \\ \hline
  $1.9136$  & $4.8508\times 10^{-2}$  & $4.3560\times 10^{-2}$  & $4.1880\times 10^{-2}$ \\ \hline
  $2.0595$  & $4.4628\times 10^{-2}$  & $3.9720\times 10^{-2}$  & $3.9140\times 10^{-2}$ \\ \hline
  $2.1975$  & $4.1058\times 10^{-2}$  & $3.7300\times 10^{-2}$  & $3.8080\times 10^{-2}$ \\ \hline
  $2.3296$  & $3.7773\times 10^{-2}$  & $3.4700\times 10^{-2}$  & $3.5680\times 10^{-2}$ \\ \hline
  $2.4570$  & $3.4751\times 10^{-2}$  & $3.3440\times 10^{-2}$  & $3.4340\times 10^{-2}$ \\ \hline
  $2.5807$  & $3.1971\times 10^{-2}$  & $3.1380\times 10^{-2}$  & $3.0680\times 10^{-2}$ \\ \hline
  $2.7012$  & $2.9413\times 10^{-2}$  & $2.9940\times 10^{-2}$  & $3.0380\times 10^{-2}$ \\ \hline
  $2.8191$  & $2.7060\times 10^{-2}$  & $2.7240\times 10^{-2}$  & $2.9240\times 10^{-2}$ \\ \hline
  $2.9348$  & $2.4895\times 10^{-2}$  & $2.5700\times 10^{-2}$  & $2.6160\times 10^{-2}$ \\ \hline
  $3.0486$  & $2.2904\times 10^{-2}$  & $2.3680\times 10^{-2}$  & $2.3200\times 10^{-2}$ \\ \hline
  $3.1608$  & $2.1071\times 10^{-2}$  & $2.2100\times 10^{-2}$  & $2.2620\times 10^{-2}$ \\ \hline
  $3.2717$  & $1.9386\times 10^{-2}$  & $2.0640\times 10^{-2}$  & $2.1780\times 10^{-2}$ \\ \hline
  $3.3813$  & $1.7835\times 10^{-2}$  & $1.9040\times 10^{-2}$  & $2.0260\times 10^{-2}$ \\ \hline
  $3.4899$  & $1.6408\times 10^{-2}$  & $1.7820\times 10^{-2}$  & $1.8560\times 10^{-2}$ \\ \hline
  $3.5976$  & $1.5095\times 10^{-2}$  & $1.6500\times 10^{-2}$  & $1.5580\times 10^{-2}$ \\ \hline
  $3.7044$  & $1.3888\times 10^{-2}$  & $1.5160\times 10^{-2}$  & $1.5120\times 10^{-2}$ \\ \hline
  $3.8106$  & $1.2777\times 10^{-2}$  & $1.4000\times 10^{-2}$  & $1.4440\times 10^{-2}$ \\ \hline
  $3.9162$  & $1.1755\times 10^{-2}$  & $1.3000\times 10^{-2}$  & $1.2280\times 10^{-2}$ \\ \hline
\end{tabular}
\end{center}
\end{table}

\addtocounter{table}{-1}
\renewcommand{\thetable}{\Roman{table}  (Cont.)}
\begin{table}[!ht]
\caption{Empirical Probability and Its Approximations ($\alpha=0.08$, $\beta=0.92$)}
\vspace{-0.5cm}
\begin{center}
\begin{tabular}{|p{2.5cm}|p{2.5cm}|p{2.5cm}|p{2.5cm}|}
\hline
Distance ($d_{i}$) & App.Theorem 4 & App.Theorem 2 & Empirical Prob. \\
\hline
\hline
  $4.0212$  & $1.0814\times 10^{-2}$  & $1.1820\times 10^{-2}$  & $1.1580\times 10^{-2}$ \\ \hline
  $4.1258$  & $9.9491\times 10^{-3}$  & $1.0840\times 10^{-2}$  & $1.1740\times 10^{-2}$ \\ \hline
  $4.2300$  & $9.1532\times 10^{-3}$  & $1.0100\times 10^{-2}$  & $1.0960\times 10^{-2}$ \\ \hline
  $4.3337$  & $8.4210\times 10^{-3}$  & $9.3800\times 10^{-3}$  & $9.5800\times 10^{-3}$ \\ \hline
  $4.4372$  & $7.7473\times 10^{-3}$  & $8.6000\times 10^{-3}$  & $8.0000\times 10^{-3}$ \\ \hline
  $4.5404$  & $7.1275\times 10^{-3}$  & $7.9600\times 10^{-3}$  & $8.2000\times 10^{-3}$ \\ \hline
  $4.6433$  & $6.5573\times 10^{-3}$  & $7.4000\times 10^{-3}$  & $7.4200\times 10^{-3}$ \\ \hline
  $4.7460$  & $6.0327\times 10^{-3}$  & $6.7800\times 10^{-3}$  & $6.1600\times 10^{-3}$ \\ \hline
  $4.8485$  & $5.5501\times 10^{-3}$  & $6.3000\times 10^{-3}$  & $6.2600\times 10^{-3}$ \\ \hline
  $4.9509$  & $5.1061\times 10^{-3}$  & $5.8600\times 10^{-3}$  & $5.9000\times 10^{-3}$ \\ \hline
  $5.0530$  & $4.6976\times 10^{-3}$  & $5.3400\times 10^{-3}$  & $5.0600\times 10^{-3}$ \\ \hline
  $5.1551$  & $4.3218\times 10^{-3}$  & $4.9400\times 10^{-3}$  & $5.2600\times 10^{-3}$ \\ \hline
  $5.2570$  & $3.9760\times 10^{-3}$  & $4.6000\times 10^{-3}$  & $4.4800\times 10^{-3}$ \\ \hline
  $5.3589$  & $3.6580\times 10^{-3}$  & $4.3000\times 10^{-3}$  & $4.3600\times 10^{-3}$ \\ \hline
  $5.4606$  & $3.3653\times 10^{-3}$  & $4.0200\times 10^{-3}$  & $3.5400\times 10^{-3}$ \\ \hline
  $5.5622$  & $3.0961\times 10^{-3}$  & $3.7000\times 10^{-3}$  & $3.4800\times 10^{-3}$ \\ \hline
  $5.6638$  & $2.8484\times 10^{-3}$  & $2.5800\times 10^{-3}$  & $1.6000\times 10^{-3}$ \\ \hline
\end{tabular}
\end{center}
\end{table}

 Consistent results have been obtained where the simulation settings such as the initial probability of the Markov chain, initial value of the PCM of the RSEIO, number of the simulated PCMs at its stationary state, etc., are changed to other values. These results suggest that both Theorem 2 and Theorem 4 can in general provide a highly accurate approximation for the stationary PCM process. Moreover, while the approximation accuracy of Theorem 4 is influenced by the parameters $\alpha$ and $\beta$ of the Markov chain $\gamma_{k}$, they do not affect that of Theorem 2. But to reach a high accuracy, the approximation of Theorem 2 usually asks for a large number of time series samples.

\section{Concluding Remarks}

In this paper, asymptotic properties of the pseudo-covariance matrix of a robust recursive state estimator are investigated under the situation that the data loss process is described by a Markov chain. It has been made clear that when the modified plant is both controllable and observable, this PCM process converges exponentially to a stationary process that does not depend on its initial value. Moreover, when this robust state estimator starts from the stabilizing solution of the algebraic Riccati equation defined by the system parameters of the modified plant, it is shown that this PCM process becomes ergodic. An important observation is that when the data arrival probability is approximately equal to $1$, the distribution of the stationary PCM process can be well approximated by a set of delta functions. Based on these results, two approximations have been derived for the stationary distribution of this PCM process, together with an error bound for one of these two approximations. Numerical simulations show that these approximations usually have a high accuracy.

As a further research, it is important to investigate characteristics of the delta functions utilized in the aforementioned approximations, as well as tighter error bounds for these approximations.

\renewcommand{\theequation}{a.\arabic{equation}}
\setcounter{equation}{0}

\section*{Appendix: Proof of Some Technical Results}

\hspace*{-0.4cm}{\bf Proof of Theorem 1:} Define $\alpha_{1h}$ and $\alpha_{0h}$ respectively as
\begin{eqnarray*}
& & \alpha_{0h}=\sup_{X,Y>0,\;X\neq Y}\frac{\delta\left({\rm\bf H}_{m}(M^{[0]},X),\;{\rm\bf H}_{m}(M^{[0]},Y)\right)}{\delta(X,Y)} \\
& & \alpha_{1h}=\sup_{X,Y>0,\;X\neq Y}\frac{\delta\left({\rm\bf H}_{m}(M^{[1]},X),\;{\rm\bf H}_{m}(M^{[1]},Y)\right)}{\delta(X,Y)}
\end{eqnarray*}
Clearly, when $\star$ belongs to the set $\{\:0,\;1\:\}$, it can be declared from these definitions that for every PDM pair $X$ and $Y$,
\begin{equation}
\delta\left({\rm\bf H}_{m}(M^{[\star]},X),\;{\rm\bf H}_{m}(M^{[\star]},Y)\right)\leq\alpha_{1h}^{\star}\alpha_{0h}^{1-\star}\delta(X,Y)
\end{equation}
On the other hand, based on the properties of a Hamiltonian matrix and Homographic transformations, it has been proved in \cite{Bougerol93,zhou14} that $\alpha_{0h}\leq 1$ is valid for all invertible $A^{[0]}$, and when  $(A^{[1]},\;G^{[1]})$ is controllable and $(H^{[1]},\;A^{[1]})$ is observable, $\alpha_{1h}<1$, provided that $A^{[1]}$ is of full rank.

Hence, for arbitrary PDMs $X$ and $Y$,
\begin{eqnarray}
\delta_{k}(X,Y)&=&\delta\left[\Phi_{k}(X),\;\Phi_{k}(Y)\right]\nonumber\\
&=&\delta\left\{{\rm\bf H}_{m}\left(\prod_{i=k}^{1}M^{[\gamma_{i}]},\;X\right),\;{\rm\bf H}_{m}\left(\prod_{i=k}^{1}M^{[\gamma_{i}]},\;Y\right)\right\}\nonumber\\
&=&\delta\left\{{\rm\bf H}_{m}\left[M^{[\gamma_{k}]},\;{\rm\bf H}_{m}\left(\prod_{i=k-1}^{1}M^{[\gamma_{i}]},\;X\right)\right],\;{\rm\bf H}_{m}\left[M^{[\gamma_{k}]},\;{\rm\bf H}_{m}\left(\prod_{i=k-1}^{1}M^{[\gamma_{i}]},\;Y\right)\right]\right\}\nonumber\\
&\leq & \alpha_{1h}^{\gamma_{k}}\alpha_{0h}^{1-\gamma_{k}}\delta\left\{{\rm\bf H}_{m}\left(\prod_{i=k-1}^{1}M^{[\gamma_{i}]},\;X\right),\;{\rm\bf H}_{m}\left(\prod_{i=k-1}^{1}M^{[\gamma_{i}]},\;Y\right)\right\}\nonumber\\
&=&\alpha_{1h}^{\gamma_{k}}\alpha_{0h}^{1-\gamma_{k}}\delta\!\left\{\!\!{\rm\bf H}_{m}\!\!\left[\!M^{[\gamma_{k-1}]},\;{\rm\bf H}_{m}\!\!\left(\prod_{i=k-2}^{1}\!\!\!M^{[\gamma_{i}]},\;X\!\right)\!\right],\;{\rm\bf H}_{m}\!\!\left[M^{[\gamma_{k-1}]},\;{\rm\bf H}_{m}\!\!\left(\prod_{i=k-2}^{1}\!\!\!M^{[\gamma_{i}]},\;Y\!\right)\!\right]\!\right\}\nonumber\\
&\leq & \alpha_{1h}^{\gamma_{k}}\alpha_{0h}^{1-\gamma_{k}}\alpha_{1h}^{\gamma_{k-1}}\alpha_{0h}^{1-\gamma_{k-1}}\delta\left\{{\rm\bf H}_{m}\left(\prod_{i=k-2}^{1}M^{[\gamma_{i}]},\;X\right),\;{\rm\bf H}_{m}\left(\prod_{i=k-2}^{1}M^{[\gamma_{i}]},\;Y\right)\right\}\nonumber\\
&=& \cdots \nonumber\\
&\leq & \left(\prod_{i=k}^{1}\alpha_{1h}^{\gamma_{i}}\alpha_{0h}^{1-\gamma_{i}}\right)\delta(X,Y)\nonumber\\
&=&\alpha_{1h}^{\sum_{i=1}^{k}\gamma_{i}}\alpha_{0h}^{k-\sum_{i=1}^{k}\gamma_{i}}\delta(X,Y)\nonumber\\
&\leq& \alpha_{1h}^{\sum_{i=1}^{k}\gamma_{i}}\delta(X,Y)
\label{eqn:a11}
\end{eqnarray}

Define a function $f(\cdot)$ on the random process $\gamma_{k}$ as $f(\gamma_{k})=\gamma_{k}$. When both $\alpha$ and $\beta$ belong to $(0,\;1)$, it is obvious that the Markov chain $\gamma_{k}$ is positive recurrent and only has two states, that is $\gamma_{k}=1$ and $\gamma_{k}=0$. Using the same symbols of Lemma 1, it is obvious that for an arbitrary $j\in\{\:0,\;1\:\}$,
\begin{equation}
f_{\alpha}^{[j]}=\sum_{k=\tau_{\alpha}^{[j]}}^{\tau_{\alpha+1}^{[j]}-1}f(\gamma_{k})\leq (\tau_{\alpha+1}^{[j]}-1)-(\tau_{\alpha}^{[j]}-1)=\tau_{\alpha+1}^{[j]}-\tau_{\alpha}^{[j]}
\end{equation}

From this relation and properties of Markov chains, it is straightforward to prove that
\begin{eqnarray}
& & s(f)=\frac{1}{\mu_{1}}>0,\hspace{0.5cm}
{\rm\bf E}\left(|f_{\alpha}^{[j]}|^{3}\right)\leq {\rm\bf E}\left(|\tau_{\alpha+1}^{[j]}-\tau_{\alpha}^{[j]}|^{3}\right)<\infty \\
& & {\rm\bf V}_{ar}\{f_{\alpha}^{[j]}-s(f)(\tau_{\alpha+1}^{[j]}-\tau_{\alpha}^{[j]})\}>0
\end{eqnarray}
Moreover, both $\mu_{1}$ and $\pi_{1}$ are positive constants. Hence, according to Lemma 1, we have that
\begin{equation}
\sup_{t\in{\cal R}}\left|{\rm\bf P}_{r}\left\{\;\frac{\sqrt{\mu_{1}}}{\sigma_{1}\sqrt{k}}\left(\sum_{i=0}^{k}\gamma_{i}-\frac{k+1}{\mu_{1}}\right)<t\right\}-\Phi(t)\right|
=O\left[\left(\frac{ln(k)}{k}\right)^{1/4}\right]
\label{eqn:a1}
\end{equation}

From this equation, it can be declared that for an arbitrary positive $\varepsilon_{1}$, there exists a positive integer $N_{1}(\varepsilon_{1})$, such that
\begin{equation}
\left|{\rm\bf P}_{r}\left\{\;\frac{\sqrt{\mu_{1}}}{\sigma_{1}\sqrt{k}}\left(\sum_{i=0}^{k}\gamma_{i}-\frac{k+1}{\mu_{1}}\right)<t\right\}-\Phi(t)\right|\leq\varepsilon_{1}
\end{equation}
is valid for every real $t$, provided that $k\geq N_{1}(\varepsilon_{1})$.

Therefore, when $k\geq N_{1}(\varepsilon_{1})$, the following relations are always valid
\begin{eqnarray}
& & -\varepsilon_{1}\leq {\rm\bf P}_{r}\left\{\;\frac{\sqrt{\mu_{1}}}{\sigma_{1}\sqrt{k}}\left(\sum_{i=0}^{k}\gamma_{i}-\frac{k+1}{\mu_{1}}\right)<-t\right\}-\Phi(-t)\leq\varepsilon_{1} \\
& & -\varepsilon_{1}\leq {\rm\bf P}_{r}\left\{\;\frac{\sqrt{\mu_{1}}}{\sigma_{1}\sqrt{k}}\left(\sum_{i=0}^{k}\gamma_{i}-\frac{k+1}{\mu_{1}}\right)<t\right\}-\Phi(t)\leq\varepsilon_{1}
\end{eqnarray}
which further leads to that for an arbitrary positive $t$,
\begin{eqnarray}
& & {\rm\bf P}_{r}\left\{\;\left|\frac{\sqrt{\mu_{1}}}{\sigma_{1}\sqrt{k}}\left(\sum_{i=0}^{k}\gamma_{i}-\frac{k+1}{\mu_{1}}\right)\right|<t\right\}\nonumber\\
&=& {\rm\bf P}_{r}\left\{\;\frac{\sqrt{\mu_{1}}}{\sigma_{1}\sqrt{k}}\left(\sum_{i=0}^{k}\gamma_{i}-\frac{k+1}{\mu_{1}}\right)<t\right\}
-{\rm\bf P}_{r}\left\{\;\frac{\sqrt{\mu_{1}}}{\sigma_{1}\sqrt{k}}\left(\sum_{i=0}^{k}\gamma_{i}-\frac{k+1}{\mu_{1}}\right)<-t\right\}\nonumber\\
&\geq & [\Phi(t)-\varepsilon_{1}]-[\Phi(-t)+\varepsilon_{1}] \nonumber\\
&=& \Phi(t)-\Phi(-t)-2\varepsilon_{1}
\label{eqn:a2}
\end{eqnarray}

On the other hand, $\left|\frac{\sqrt{\mu_{1}}}{\sigma_{1}\sqrt{k}}\left(\sum_{i=0}^{k}\gamma_{i}-\frac{k+1}{\mu_{1}}\right)\right|<t$ is equivalent to
\begin{equation}
\left|\sum_{i=0}^{k}\gamma_{i}-(k+1)\pi_{1}\right|<\sigma_{1} t\sqrt{\pi_{1}}\sqrt{k}
\end{equation}
which implies that
\begin{eqnarray}
\sum_{i=0}^{k}\gamma_{i}&>&(k+1)\pi_{1}-\sigma_{1} t\sqrt{\pi_{1}}\sqrt{k}\nonumber\\
&=&\sqrt{k}\left(\frac{k+1}{\sqrt{k}}\sqrt{\pi_{1}}-\sigma_{1}t\right)\sqrt{\pi_{1}}
\end{eqnarray}

Note that $\pi_{1}>0$ and is independent of $k$. It is obvious that $\frac{k+1}{\sqrt{k}}\sqrt{\pi_{1}}-\sigma_{1}t$ is a monotonically increasing function of $k$. This means that for an arbitrary positive $t$, there exists a positive integer $N_{2}(t)$, such that $\frac{k+1}{\sqrt{k}}\sqrt{\pi_{1}}-\sigma_{1}t>0$ is valid for each $k\geq N_{2}(t)$.

Define $N_{2}(t)$ and $\xi(t)$ respectively as
\begin{eqnarray*}
& & N_{2}(t)=\min\left\{\:k\:\left|\: k\;{\rm is\;\; an \;\; integer},\;\; \frac{k+1}{\sqrt{k}}\sqrt{\pi_{1}}-\sigma_{1}t>0\:\right.\right\} \\
& & \xi(t)=\left(\frac{N_{2}(t)+2}{\sqrt{N_{2}(t)+1}}\sqrt{\pi_{1}}-\sigma_{1}t\right)\sqrt{\pi_{1}}
\end{eqnarray*}
Then, it is obvious that for an arbitrary $k\geq N_{2}(t)+1$,
$\left(\frac{k+1}{\sqrt{k}}\sqrt{\pi_{1}}-\sigma_{1}t\right)\sqrt{\pi_{1}}\geq \xi(t)>0$, which further leads to
\begin{equation}
\sum_{i=0}^{k}\gamma_{i}>\sqrt{k}\xi(t)
\label{eqn:a3}
\end{equation}

In addition, from the definition of the function $\Phi(t)$ or the properties of the normal distribution with mathematical expectation and variance respectively being $0$ and $1$, it can be declared that for an arbitrary $\varepsilon_{2}>0$, there exists a positive $t(\varepsilon_{2})$, such that
\begin{equation}
\Phi[t(\varepsilon_{2})]-\Phi[-t(\varepsilon_{2})]\geq 1-\varepsilon_{2}
\label{eqn:a4}
\end{equation}

Now, for an arbitrary positive $\varepsilon$, let $\varepsilon_{1}=\frac{\varepsilon}{4}$ and $\varepsilon_{2}=\frac{\varepsilon}{2}$. Define $N(\varepsilon)$ as $N(\varepsilon)=\max\{N_{1}(\varepsilon_{1}),\; N_{2}(t(\varepsilon_{2}))$ $+1\}$. Then, from Equations (\ref{eqn:a2}) and (\ref{eqn:a4}), we have that when $k$ is larger than $N(\varepsilon)$, it is certain that
\begin{equation}
{\rm\bf P}_{r}\left\{\;\left|\frac{\sqrt{\mu_{1}}}{\sigma_{1}\sqrt{k}}\left(\sum_{i=0}^{k}\gamma_{i}-\frac{k+1}{\mu_{1}}\right)\right|<t(\varepsilon_{2})\right\}
\geq 1-\frac{\varepsilon}{2}-2\times\frac{\varepsilon}{4}=1-\varepsilon
\end{equation}

Based on this relation and Equation (\ref{eqn:a3}), it can be further declared that
\begin{equation}
{\rm\bf P}_{r}\left\{\;\sum_{i=0}^{k}\gamma_{i}>\sqrt{k}\xi\left[t\left(\frac{\varepsilon}{2}\right)\right]\right\}
\geq
{\rm\bf P}_{r}\left\{\;\left|\frac{\sqrt{\mu_{1}}}{\sigma_{1}\sqrt{k}}\left(\sum_{i=0}^{k}\gamma_{i}-\frac{k+1}{\mu_{1}}\right)\right|<t(\varepsilon_{2})\right\}
\geq 1-\varepsilon
\end{equation}

A combination of this inequality and Equation (\ref{eqn:a11}) makes it clear that if $k\geq N(\varepsilon)$, then, with a probability greater than $1-\varepsilon$,
\begin{equation}
\delta_{k}(X,Y)\leq \alpha_{1h}^{\sqrt{k}\xi[t(\varepsilon/2)]}\delta(X,Y)
\label{eqn:a10}
\end{equation}

As $0\leq \alpha_{1h}<1$ and $\delta(X,Y)$ is a finite positive number when both $X$ and $Y$ are finite PDMs, it can therefore be declared that $\lim_{k\rightarrow\infty}\alpha_{1h}^{\sqrt{k}\xi[t(\varepsilon/2)]}=0$. Recall that $\delta_{k}(X,Y)$ is nonnegative and $\varepsilon$ is an arbitrarily selected positive number, these relations mean that for any finite PDMs $X$ and $Y$, $\lim_{k\rightarrow\infty}\delta_{k}(X,Y)=0$ in probability.
This completes the proof. \hspace{\fill}$\Diamond$

\hspace*{-0.4cm}{\bf Proof of Theorem 2:} Assume that $P^{[j]}=\lim_{k\rightarrow\infty}{\rm\bf H}_{m}\left(M^{[\gamma_{0}^{[j]}]}
M^{[\gamma_{-1}^{[j]}]}\cdots M^{[\gamma_{-k}^{[j]}]},\;P^{\star}\right)$. Then, for each $\#\in\{\:0,\;1\:\}$,
\begin{eqnarray}
{\rm\bf H}_{m}\left(M^{[\#]},P^{[j]}\right)&=& {\rm\bf H}_{m}\left(M^{[\#]},\lim_{k\rightarrow\infty}{\rm\bf H}_{m}\left(M^{[\gamma_{0}^{[j]}]}
M^{[\gamma_{-1}^{[j]}]}\cdots M^{[\gamma_{-k}^{[j]}]},\;P^{\star}\right)\right)\nonumber\\
&=&{\rm\bf H}_{m}\left(M^{[\#]}\lim_{k\rightarrow\infty}M^{[\gamma_{0}^{[j]}]}
M^{[\gamma_{-1}^{[j]}]}\cdots M^{[\gamma_{-k}^{[j]}]},\;P^{\star}\right)\nonumber\\
&=& \lim_{k\rightarrow\infty}{\rm\bf H}_{m}\left(M^{[\#]}M^{[\gamma_{0}^{[j]}]}
M^{[\gamma_{-1}^{[j]}]}\cdots M^{[\gamma_{-k}^{[j]}]},\;P^{\star}\right)\nonumber\\
&=& \lim_{k\rightarrow\infty}{\rm\bf H}_{m}\left(M^{[\#]}M^{[\gamma_{0}^{[j]}]}
M^{[\gamma_{-1}^{[j]}]}\cdots M^{[\gamma_{-k+1}^{[j]}]},\;{\rm\bf H}_{m}\left(M^{[\gamma_{-k}^{[j]}]},\;P^{\star}\right)\right)\nonumber\\
&=& \lim_{k\rightarrow\infty}{\rm\bf H}_{m}\left(M^{[\#]}M^{[\gamma_{0}^{[j]}]}
M^{[\gamma_{-1}^{[j]}]}\cdots M^{[\gamma_{-k+1}^{[j]}]},\;P^{\star}\right)\nonumber\\
&=& P^{[\bar{j}]}
\label{eqn:a5}
\end{eqnarray}
in which $j=n(\gamma_{i}^{[j]}|_{i=0}^{-\infty})=\sum_{i=0}^{-\infty}\gamma_{i}^{[j]}2^{i}$,
$\bar{j}=n(\bar{\gamma}_{i}^{[j]}|_{i=0}^{-\infty})=\sum_{i=0}^{-\infty}\bar{\gamma}_{i}^{[j]}2^{i}$, and $\bar{\gamma}_{i}^{[j]}={\gamma}_{i+1}^{[j]}$ whenever $i\leq -1$ while $\bar{\gamma}_{0}^{[j]}=\#$.

Define $n_{in}(\#,\gamma_{i}^{[j]}|_{i=0}^{-\infty})$ as $n_{in}(\#,\gamma_{i}^{[j]}|_{i=0}^{-\infty})=\bar{j}-j$. Then
\begin{eqnarray}
n_{in}(\#,\gamma_{i}^{[j]}|_{i=0}^{-\infty})&=&\sum_{i=0}^{-\infty}\bar{\gamma}_{i}^{[j]}2^{i}-\sum_{i=0}^{-\infty}\gamma_{i}^{[j]}2^{i}\nonumber\\
&=&(\#-\gamma_{0}^{[j]})+\sum_{i=-1}^{-\infty}({\gamma}_{i+1}^{[j]}-{\gamma}_{i}^{[j]})2^{i}
\end{eqnarray}

For a given sequence $\gamma_{l}^{[j]}|_{l=1}^{s}$ with $\gamma_{l}^{[j]}\in\{0,\;1\}$, define $\gamma_{i,l}^{[j]}$ as $\gamma_{i,0}^{[j]}=\gamma_{i}^{[j]}$, $i=0,-1,\cdots$, and
\begin{equation}
\gamma_{i,l}^{[j]}=\left\{\begin{array}{lll}
\gamma_{i+1,l-1}^{[j]} & \;\; & i\leq -1 \\ \gamma_{l}^{[j]} &\;\; & i=0 \end{array}\right.\hspace{1cm} l=1,2,\cdots,s
\end{equation}
Denote ${\rm\bf H}_{m}\left(M^{[\gamma_{l}^{[j]}]}M^{[\gamma_{l-1}^{[j]}]}\cdots M^{[\gamma_{1}^{[j]}]},P^{[j]}\right)$ by $P^{[j_{l}]}$, $l=1,2,\cdots,s$. Then, a repetitive utilization of Equation (\ref{eqn:a5}) leads to
\begin{eqnarray}
j_{s}&=& j_{s-1}+n_{in}(\gamma_{s}^{[j]},\gamma_{i,s-1}^{[j]}|_{i=0}^{-\infty}) \nonumber\\
&=& j_{s-2}+n_{in}(\gamma_{s-1}^{[j]},\gamma_{i,s-2}^{[j]}|_{i=0}^{-\infty})+n_{in}(\gamma_{s}^{[j]},\gamma_{i,s-1}^{[j]}|_{i=0}^{-\infty})\nonumber\\
&=&\cdots \nonumber\\
&=& j_{0}+\sum_{l=1}^{s}n_{in}(\gamma_{l}^{[j]},\gamma_{i,l-1}^{[j]}|_{i=0}^{-\infty})
\end{eqnarray}
in which $j_{0}=j$. Therefore, $j_{s}=j$ if and only if
\begin{equation}
\sum_{l=1}^{s}n_{in}(\gamma_{l}^{[j]},\gamma_{i,l-1}^{[j]}|_{i=0}^{-\infty})=0
\end{equation}

On the other hand, from the definition of $n_{in}(\#,\gamma_{i}^{[j]}|_{i=0}^{-\infty})$, straightforward algebraic manipulations show that
\begin{eqnarray}
\sum_{l=1}^{s}n_{in}(\gamma_{l}^{[j]},\gamma_{i,l-1}^{[j]}|_{i=0}^{-\infty})&=&
\sum_{l=1}^{s}\left[(\gamma_{l}^{[j]}-\gamma_{0,l-1}^{[j]})+\sum_{i=-1}^{-\infty}(\gamma_{i+1,l-1}^{[j]}-\gamma_{i,l-1}^{[j]})2^{i}\right]\nonumber\\
&=&\sum_{l=1}^{s}(\gamma_{l}^{[j]}-\gamma_{0,l-s}^{[j]})2^{l-s}+\sum_{i=0}^{-\infty}(\gamma_{0,i}^{[j]}-\gamma_{0,i-s}^{[j]})2^{i-s}
\end{eqnarray}

Therefore, $j_{s}=j$ if and only if
\begin{eqnarray}
\gamma_{s}^{[j]}&=&\gamma_{0,0}^{[j]}-\sum_{l=1}^{s-1}(\gamma_{l}^{[j]}-\gamma_{0,l-s}^{[j]})2^{l-s}-\sum_{i=0}^{-\infty}(\gamma_{0,i}^{[j]}-\gamma_{0,i-s}^{[j]})2^{i-s}\nonumber\\
&=&\gamma_{0,0}^{[j]}+\sum_{l=1}^{s-1}(\gamma_{0,l-s}^{[j]}-\gamma_{l}^{[j]})2^{l-s}+\sum_{i=0}^{-\infty}(\gamma_{0,i-s}^{[j]}-\gamma_{0,i}^{[j]})2^{i-s}\nonumber\\
&=&\gamma_{0,0}^{[j]}+\sum_{i=-1}^{1-s}\gamma_{0,i}^{[j]}2^{i}+2^{-s}\sum_{i=0}^{-\infty}(\gamma_{0,i-s}^{[j]}-\gamma_{0,i}^{[j]})2^{i}
-\sum_{i=1}^{s-1}\gamma_{s-i}^{[j]}2^{-i}
\end{eqnarray}

Define a set ${\cal S}^{[j]}$ as
\begin{displaymath}
{\cal S}^{[j]}=\left\{\:k\:\left|\:k=\min (s), \begin{array}{l} \gamma_{0,0}^{[j]}+\sum_{i=-1}^{1-s}\gamma_{0,i}^{[j]}2^{i}+2^{-s}\sum_{i=0}^{-\infty}(\gamma_{0,i-s}^{[j]}-\gamma_{0,i}^{[j]})2^{i}
-\sum_{i=1}^{s-1}\gamma_{s-i}^{[j]}2^{-i}\in\{0,\;1\} \\
\gamma_{i}^{[j]}\in\{0,\;1\},\;i=1,2,\cdots,s-1\end{array} \:\right.\right\}
\end{displaymath}
Assume that the set ${\cal S}^{[j]}$ is not empty for all the possible $j$.
Then, for any $s\in{\cal S}^{[j]}$, there exists a binary sequence $\gamma_{i}^{[j]}|_{i=1}^{s}$ with $\gamma_{i}^{[j]}\in\{0,\;1\}$ such that
\begin{equation}
{\rm\bf H}_{m}\left(M^{[\gamma_{s}^{[j]}]}M^{[\gamma_{s-1}^{[j]}]}\cdots M^{[\gamma_{1}^{[j]}]},P^{[j]}\right)=P^{[j]}
\end{equation}

Assume that the Markov chain $\gamma_{k}$ is in its stationary state in which both ${\rm\bf P}_{r}(\gamma_{k}=1)$ and ${\rm\bf P}_{r}(\gamma_{k}=0)$ take a constant value belonging to $(0,\;1)$. Denote $\max\{{\rm\bf P}_{r}(\gamma_{k}=1),\;{\rm\bf P}_{r}(\gamma_{k}=0)\}$ and $\min\{{\rm\bf P}_{r}(\gamma_{k}=1),\;{\rm\bf P}_{r}(\gamma_{k}=0)\}$ respectively by $p_{hs}$ and $p_{ls}$. Moreover, for a particular $s\in{\cal S}^{[j]}$, denote the corresponding $\gamma_{i}^{[j]}|_{i=1}^{s}$ by $\gamma_{i}^{[j,s]}|_{i=1}^{s}$. Then,
\begin{equation}
{\rm\bf P}_{r}(s)=\prod_{i=1}^{s}\left[\gamma_{i}^{[j,s]}{\rm\bf P}_{r}(\gamma_{i}^{[j,s]}=1)+(1-\gamma_{i}^{[j,s]}){\rm\bf P}_{r}(\gamma_{i}^{[j,s]}=0)\right]
\end{equation}
Therefore
\begin{eqnarray}
& & {\rm\bf P}_{r}(s)\geq \prod_{i=1}^{s}\min\{{\rm\bf P}_{r}(\gamma_{k}=1),\;{\rm\bf P}_{r}(\gamma_{k}=0)\}=p_{ls}^{s} \\
& & {\rm\bf P}_{r}(s)\leq \prod_{i=1}^{s}\max\{{\rm\bf P}_{r}(\gamma_{k}=1),\;{\rm\bf P}_{r}(\gamma_{k}=0)\}=p_{hs}^{s} \\
\end{eqnarray}

Hence, when an integer $s$ belonging to the set ${\cal S}^{[j]}$ takes a finite value, its occurrence probability is certainly greater than $0$.

As in Lemma 1, let $\tau_{v}^{[j]}$ denote the $v$-th time instant that $j_{s}=j_{0}$ and $f_{v}^{[j]}(P_{k|k})$ the random variable $\sum_{k=\tau_{v}^{[j]}}^{\tau_{v+1}^{[j]}-1}I_{{\cal B}_{\varepsilon}}(P_{k|k})$. Then,
\begin{equation}
\left|f_{v}^{[j]}(P_{k|k})\right|=\left|\sum_{k=\tau_{v}^{[j]}}^{\tau_{v+1}^{[j]}-1}I_{{\cal B}_{\varepsilon}}(P_{k|k})\right|\leq
(\tau_{v+1}^{[j]}-1)-(\tau_{v}^{[j]}-1)=\tau_{v+1}^{[j]}-\tau_{v}^{[j]}\in{\cal S}^{[j]}
\end{equation}
Hence
\begin{equation}
{\rm\bf E}\left\{\left|f_{v}^{[j]}(P_{k|k})\right|^{3}\right\}\leq {\rm\bf E}\left\{(\tau_{v+1}^{[j]}-\tau_{v}^{[j]})^{3}\right\}
=\sum_{k\in{\cal S}^{[j]}}k^{3}{\rm\bf P}_{r}(k)
\leq \sum_{k=1}^{\infty}k^{3}p_{hs}^{k}
\end{equation}

Note that $k^{3}=(k+1)k(k-1)+k$. It can be directly proved that
\begin{equation}
\sum_{k=1}^{\infty}k^{3}p_{hs}^{k}=\frac{(1+p_{hs})^{2}+2p_{hs}}{(1-p_{hs})^{4}}p_{hs}
\end{equation}
Therefore, when $p_{hs}$ belongs to $(0,\;1)$, both ${\rm\bf E}\left\{\left|f_{v}^{[j]}(P_{k|k})\right|^{3}\right\}$ and ${\rm\bf E}\left\{(\tau_{v+1}^{[j]}-\tau_{v}^{[j]})^{3}\right\}$ are finite.

Note also that ${\rm\bf H}_{m}\left(M^{[1]},P^{\star}\right)=P^{\star}$ and $\lim_{k\rightarrow\infty}{\rm\bf H}_{m}\left[M^{[1]k},{\rm\bf H}_{m}\left(M^{[0]},P^{\star}\right)\right]=P^{\star}$ in probability. It is obvious that when $j=\sum_{i=0}^{-\infty}2^{i}$, the set ${\cal S}^{[j]}$ has at least two finite integers that has an occurrence probability greater than $0$. Therefore, when the PCM of RSEIO is started from $P^{\star}$, the corresponding $f_{\alpha}^{[j]}-s(f)(\tau_{\alpha+1}^{[j]}-\tau_{\alpha}^{[j]})$ has a variance greater than $0$.

Denote $\sum_{k\in{\cal S}^{[j]}}k {\rm\bf P}_{r}(k)$ by $\mu^{[j]}$. Then, it can be directly proved that
\begin{equation}
F(\varepsilon)=\sum_{j\in {\cal I}}\frac{1}{\mu^{[j]}}I_{{\cal B}_{\varepsilon}}(P^{[j]})
\end{equation}

On the other hand, according to Lemma 1, we have that
\begin{equation}
\lim_{n\rightarrow\infty}\frac{1}{n+1}\sum_{k=0}^{n}I_{{\cal B}_{\varepsilon}}(P_{k|k})=\sum_{j\in {\cal I}}\frac{1}{\mu^{[j]}}I_{{\cal B}_{\varepsilon}}(P^{[j]})
\end{equation}
and the convergence rate is of the order $\left(\frac{ln(n)}{n}\right)^{1/4}$. The proof can now be completed for the case in which ${\cal S}^{[j]}\neq \emptyset$ for every possible $j$, through combining the above two equations together.

If there exists a $j$ such that the set ${\cal S}^{[j]}$ is empty, the conclusions can still be established through modifying $P^{[j]}$ to ${\cal P}^{[j]}(\varepsilon_{s})$ in the above arguments, in which $\varepsilon_{s}$ is a prescribed positive number. More precisely, according to Theorem 1, for arbitrary $j_{1}$ and $j_{2}$, there always exists a finite step transformation from an element of ${\cal P}^{[j_{1}]}(\varepsilon_{s})$ to the set ${\cal P}^{[j_{2}]}(\varepsilon_{s})$. Therefore, the corresponding set ${\cal S}^{[j]}$ is certainly not empty. The results can then be established through decreasing $\varepsilon_{s}$ to $0$.
\hspace{\fill}$\Diamond$

\hspace*{-0.4cm}{\bf Proof of Lemma 4:} From $P_{0|0}=P^{\star}$ and ${\rm\bf H}_{m}\left(M^{[1]},\;P^{\star}\right)=P^{\star}$, it is clear that $P_{1|1}$ has only one additional possible value, that is, ${\rm\bf H}_{m}\left(M^{[0]},\;P^{\star}\right)$. Hence, the number of elements in ${\cal P}^{[1]}$ is $2$ and ${\cal P}^{[1]}=\{\:P^{\star},\;{\rm\bf H}_{m}\left(M^{[0]},\;P^{\star}\right)\:\}$.

Assume that the conclusions are valid with $n=l$. That is, $\#\left({\cal P}^{[l]}\right)=2^{l}$ and
\begin{equation}
{\cal P}^{[l]}=\left\{P^{\star},\;{\rm\bf H}_{m}\left(M^{[0]},\;P^{\star}\right)\right\}\bigcup\left\{\:P\:\left|\:\begin{array}{l}
P={\rm\bf H}_{m}\left[M^{[\gamma_{k}]}M^{[\gamma_{k-1}]}\cdots M^{[\gamma_{2}]},\;{\rm\bf H}_{m}\left(M^{[0]},\;P^{\star}\right)\right]\\
\hspace*{1cm}\gamma_{j}\in\{0,\;1\},\;j\in\{2,3,\cdots,k\},\;k\in\{2,3,\cdots,l\}\end{array}\:\right.\right\}
\end{equation}

Then, when $n=l+1$, we have
\begin{eqnarray}
{\cal P}^{[l+1]}&=&{\cal P}^{[l]}\bigcup\left\{\:P_{l+1}\:\left|\:
P_{l+1}={\rm\bf H}_{m}\left(M^{[\gamma_{l+1}]},\;P\right),\;P\in{\cal P}^{[l]}\backslash {\cal P}^{[l-1]},\;\gamma_{l+1}\in\{0,\;1\}\:\right.\right\}\nonumber\\
&=& {\cal P}^{[l]}\bigcup\left\{\:P_{l+1}\:\left|\:\begin{array}{l}
P_{l+1}={\rm\bf H}_{m}\left\{M^{[\gamma_{l+1}]},\;{\rm\bf H}_{m}\left[M^{[\gamma_{l}]}M^{[\gamma_{l-1}]}\cdots M^{[\gamma_{2}]},\;{\rm\bf H}_{m}\left(M^{[0]},\;P^{\star}\right)\right]\right\}\\
\hspace*{1cm} \gamma_{j}\in\{0,\;1\},\;j\in\{2,3,\cdots,l+1\}\end{array}\:\right.\right\}\nonumber\\
&=& {\cal P}^{[l]}\bigcup\left\{\:P_{l+1}\:\left|\:\begin{array}{l}
P_{l+1}={\rm\bf H}_{m}\left[M^{[\gamma_{l+1}]}M^{[\gamma_{l}]}\cdots M^{[\gamma_{2}]},\;{\rm\bf H}_{m}\left(M^{[0]},\;P^{\star}\right)\right] \\
\hspace*{1cm} \gamma_{j}\in\{0,\;1\},\;j\in\{2,3,\cdots,l+1\}\end{array}\:\right.\right\}
\end{eqnarray}

From the regularity of the matrices $A^{[0]}$ and $A^{[1]}$, as well as Lemma 3, it can be proved straightforwardly that
\begin{equation}
{\cal P}^{[l]}\bigcap\left\{\:P_{l+1}\:\left|\:
\begin{array}{l} P_{l+1}={\rm\bf H}_{m}\left[M^{[\gamma_{l+1}]}M^{[\gamma_{l}]}\cdots M^{[\gamma_{2}]},\;{\rm\bf H}_{m}\left(M^{[0]},\;P^{\star}\right)\right]\\
\hspace*{1cm}\gamma_{j}\in\{0,\;1\},\;j\in\{2,3,\cdots,l+1\}\end{array}\:\right.\right\}=\emptyset
\end{equation}
Therefore,
\begin{eqnarray}
& &\hspace*{-1.2cm} \#({\cal P}^{[l+1]})=\#({\cal P}^{[l]})+2^{l-1}=2^{l} \\
& & \hspace*{-1.2cm}{\cal P}^{[l+1]}\!=\!\left\{\!P^{\star},\;{\rm\bf H}_{m}\left(\!M^{[0]},\;P^{\star}\!\right)\!\right\}\bigcup\left\{\!\!P\left|\begin{array}{l}
P\!=\!{\rm\bf H}_{m}\!\!\left[\!M^{[\gamma_{k}]}M^{[\gamma_{k-1}]}\cdots M^{[\gamma_{2}]},\;{\rm\bf H}_{m}\!\left(\!M^{[0]},\;P^{\star}\!\right)\!\right]\\
\hspace*{0.0cm}\gamma_{j}\in\{0,\;1\},\;j\in\{2,3,\cdots,k\},\;k\in\{2,3,\cdots,l+1\}\end{array}\!\!\!\!\!\!\right.\right\}
\end{eqnarray}

This completes the proof. \hspace{\fill}$\Diamond$

\hspace*{-0.4cm}{\bf Proof of Theorem 3:} At first, probabilities are investigated for the occurrence of $P_{k|k}=\bar{P}^{[j]}$ with $j=1+2^{s-1}+\sum_{l=1}^{s-1}\gamma_{l}^{[j]}2^{l-1}$. From the definition of $\bar{P}^{[j]}$, it is obvious that $P_{k|k}=\bar{P}^{[j]}$ if and only if $k\geq s+1$, $\gamma_{l}=1$ when $l\in\{1,2,\cdots,k-s-1\}$, $\gamma_{k-s}=0$ and $\gamma_{i+k-s}=\gamma_{i}^{[j]}$ when $i\in\{1,2,\cdots,s\}$.

Hence,
\begin{eqnarray}
{\rm\bf P}_{r}\left(P_{k|k}=\bar{P}^{[j]}\right)&=&{\rm\bf P}_{r}\left(\gamma_{1}=1,\;\cdots,\;\gamma_{k-s-1}=1,\;\gamma_{k-s}=0,\;\gamma_{k-s+1}=\gamma_{1}^{[j]},\cdots,\gamma_{k}=\gamma_{s}^{[j]}\right)\nonumber\\
&=&\prod_{l=1}^{k-s-1}{\rm\bf P}_{r}\left(\gamma_{l}=1\right)\times {\rm\bf P}_{r}\left(\gamma_{k-s}=0\right)\prod_{i=1}^{s}{\rm\bf P}_{r}\left(\gamma_{i+k-s}=\gamma_{i}^{[j]}\right)\nonumber\\
&=& \gamma_{st}^{k-s-1}(1-\gamma_{st})p_{j}
\end{eqnarray}
in which $p_{j}=\prod_{i=1}^{s}{\rm\bf P}_{r}\left(\gamma_{i+k-s}=\gamma_{i}^{[j]}\right)$.

Therefore, the occurrence of $\bar{P}^{[j]}$ in the PCM samples $P_{0|0}$, $P_{1|1}$, $\cdots$, $P_{n|n}$ has the following probability $\bar{p}_{j}$,
\begin{equation}
\bar{p}_{j}=\sum_{k=s+1}^{n}{\rm\bf P}_{r}\left(P_{k|k}=\bar{P}^{[j]}\right)=\sum_{k=s+1}^{n}\gamma_{st}^{k-s-1}(1-\gamma_{st})p_{j}=(1-\gamma_{st}^{n-s})p_{j}
\label{eqn:a6}
\end{equation}

Note that when $\gamma_{l}^{[j]}\in\{0,\;1\}$, $l=1,2,\cdots,s$, it is certain that $0\leq\sum_{l=1}^{s-1}\gamma_{l}^{[j]}2^{l-1}\leq 2^{s-1}-1$. We therefore have that
\begin{equation}
1+2^{s-1}\leq j\leq 2^{s}
\end{equation}
which is equivalent to $1+log_{2}(j-1)\geq s\geq log_{2}(j)$. As $s$ is a positive integer, it is obvious that
\begin{equation}
s=\lceil log_{2}(j)\rceil
\label{eqn:a7}
\end{equation}
Therefore, $\gamma_{l}^{[j]}$ with $l\in\{1,2,\cdots,\lceil log_{2}(j)\rceil\}$ is the binary code of $j-1-2^{\lceil log_{2}(j)\rceil -1}$. It can therefore be declared that for any given $j$ belonging to $\{1,2,\cdots,2^{n}\}$, both $s$ and $\gamma_{l}^{[j]}|_{l=1}^{s}$ are uniquely determined through the requirement that
$j=1+2^{s-1}+\sum_{l=1}^{s-1}\gamma_{l}^{[j]}2^{l-1}$.

On the other hand, let $N_{0}(j)$ denote the number of zeros in the sequence $\gamma_{i}^{[j]}|_{i=1}^{s}$. Then,
\begin{eqnarray}
p_{j}&=& \prod_{i=1}^{s}{\rm\bf P}_{r}\left(\gamma_{i+k-s}=\gamma_{i}^{[j]}\right) \nonumber\\
&=&\prod_{i=1}^{s}\left[\gamma_{i}^{[j]}{\rm\bf P}_{r}(\gamma_{i}^{[j]}=1)+(1-\gamma_{i}^{[j]}){\rm\bf P}_{r}(\gamma_{i}^{[j]}=0)\right]\nonumber\\
&=&\prod_{i=1}^{s}{\rm\bf P}_{r}^{\gamma_{i}^{[j]}}(\gamma_{i}^{[j]}=1){\rm\bf P}_{r}^{(1-\gamma_{i}^{[j]})}(\gamma_{i}^{[j]}=0)\nonumber\\
&=&\gamma_{st}^{\sum_{i=1}^{s}\gamma_{i}^{[j]}}(1-\gamma_{st})^{s-\sum_{i=1}^{s}\gamma_{i}^{[j]}}\nonumber\\
&=&(1-\gamma_{st})^{N_{0}(j)}\gamma_{st}^{\lceil log_{2}(j)\rceil-N_{0}(j)}
\label{eqn:a8}
\end{eqnarray}

Summarizing Equations (\ref{eqn:a6}), (\ref{eqn:a7}) and (\ref{eqn:a8}), the following formula is obtained for $\bar{p}_{j}$
\begin{equation}
\bar{p}_{j}=(1-\gamma_{st}^{n-\lceil log_{2}(j)\rceil})\gamma_{st}^{\lceil log_{2}(j)\rceil-N_{0}(j)}(1-\gamma_{st})^{N_{0}(j)}
\end{equation}
Note that $N_{0}(j)=s-\sum_{i=1}^{s}\gamma_{i}^{[j]}$.
Hence, from Equation (\ref{eqn:a7}), the ergodicity of the random process $P_{k|k}$ established in Corollary 1, and the Bernoulli's law of large number \cite{mt93}, it can be claimed that
\begin{eqnarray}
& & \lim_{n\rightarrow\infty}\frac{1}{n+1}\sum_{k=0}^{n}I_{{\cal B}_{\varepsilon}}(P_{k|k})\nonumber\\
&=&\lim_{n\rightarrow\infty}\sum_{j\in{\cal N}_{\varepsilon}}\bar{p}_{j}\nonumber\\
&=&\lim_{n\rightarrow\infty}\sum_{j\in{\cal N}_{\varepsilon}}\left(1-\gamma_{st}^{n-\lceil log_{2}(j) \rceil}\right)\gamma_{st}^{\sum_{i=1}^{\lceil log_{2}(j) \rceil}\gamma_{i}^{[j]}}(1-\gamma_{st})^{\lceil log_{2}(j) \rceil-\sum_{i=1}^{\lceil log_{2}(j) \rceil}\gamma_{i}^{[j]}}
\end{eqnarray}
and the convergence rate is exponential. This completes the proof. \hspace{\fill}$\Diamond$

\hspace*{-0.4cm}{\bf Proof of Theorem 4:}
Note that ${\rm\bf H}_{m}\left(I,\; P^{\star}\right)=P^{\star}$. When the assumption is satisfied, assume that
${\rm\bf P}_{r}\left\{P_{\infty|\infty}={\rm\bf H}_{m}\left(M^{[0]i},\; P^{\star}\right)\right\}=a_{i}$, $i=0,1,\cdots$. Then, from the definition of probabilities, we have that
\begin{equation}
\sum_{i=0}^{\infty}a_{i}=1
\label{eqn:a9}
\end{equation}

On the other hand, note that ${\rm\bf H}_{m}\left[M^{[0]},\;{\rm\bf H}_{m}\left(M^{[0]i},\; P^{\star}\right)\right] ={\rm\bf H}_{m}\left(M^{[0](i+1)},\; P^{\star}\right)$. Moreover, when the Markov chain achieves its stationary state, ${\rm\bf P}_{r}(\gamma_{k}=1)=\gamma_{st}$. It can therefore be declared that when the random process $P_{k|k}$ reaches its stationary state,
\begin{equation}
{\rm\bf P}_{r}\left\{P_{n+1|n+1}={\rm\bf H}_{m}\left(M^{[0](i+1)},\; P^{\star}\right)\right\}=(1-\gamma_{st}) {\rm\bf P}_{r}\left\{P_{n|n}={\rm\bf H}_{m}\left(M^{[0]i},\; P^{\star}\right)\right\}
\end{equation}
Moreover, to guarantee the stationarity of the random process, it is necessary that
\begin{equation}
\lim_{n\rightarrow\infty} {\rm\bf P}_{r}\left\{P_{n+1|n+1}={\rm\bf H}_{m}\left(M^{[0]i},\; P^{\star}\right)\right\}=\lim_{n\rightarrow\infty} {\rm\bf P}_{r}\left\{P_{n|n}={\rm\bf H}_{m}\left(M^{[0]i},\; P^{\star}\right)\right\}
\end{equation}
Therefore,
\begin{equation}
a_{i+1}=(1-\gamma_{st})a_{i-1},\hspace{0.5cm}i=1,2,\cdots
\end{equation}

Substitute this relation into Equation (\ref{eqn:a9}), the following equation is obtained
\begin{equation}
a_{0}+(1-\gamma_{st})a_{0}+(1-\gamma_{st})^{2}a_{0}+\cdots=1
\end{equation}
Hence
\begin{equation}
a_{0}=\frac{1}{\sum_{i=0}^{\infty}(1-\gamma_{st})^{i}}=\gamma
\end{equation}
which further leads to
\begin{equation}
{\rm\bf P}_{r}\left\{P_{\infty|\infty}={\rm\bf H}_{m}\left(M^{[0]i},\; P^{\star}\right)\right\}=a_{0}(1-\gamma_{st})^{i}=\gamma_{st}(1-\gamma_{st})^{i}
\end{equation}

This completes the proof. \hspace{\fill}$\Diamond$

\end{document}